\documentclass[aps,reprint,onecolumn,amsmath,amssymb,longbibliography,notitlepage,raggedbottom]{revtex4-1}

\usepackage{graphicx}
\usepackage[caption=false]{subfig}
\usepackage{bm}
\usepackage{stmaryrd}
\usepackage{geometry} % manually change paper formatting to resemble PRF
\usepackage{color}

\begin{document}

\preprint{APS/123-QED}

\title{On non-normality and classification of amplification mechanisms in stability and resolvent analysis}% Force line breaks with \\

\author{Sean Symon}
 \email{ssymon@caltech.edu}

%\affiliation{%
 %Graduate Aerospace Laboratories, California Institute of Technology, Pasadena, California 91125, USA\\
%}%

\author{Kevin Rosenberg}

%\affiliation{%
% Graduate Aerospace Laboratories, California Institute of Technology, Pasadena, California 91125, USA\\
%}%

\author{Scott T. M. Dawson}
%\affiliation{%
% Graduate Aerospace Laboratories, California Institute of Technology, Pasadena, California 91125, USA\\
%}%

\author{Beverley J. McKeon}

\affiliation{Graduate Aerospace Laboratories, California Institute of Technology, Pasadena, California 91125, USA}

\date{\today}% It is always \today, today,

\begin{abstract}

\vspace{0.5in}

We seek to quantify non-normality of the most amplified resolvent modes and predict their features based on the characteristics of the base or mean velocity profile. A 2-by-2 model linear Navier-Stokes (LNS) operator illustrates how non-normality from mean shear distributes perturbation energy in different velocity components of the forcing and response modes. The inverse of their inner product, which is unity for a purely normal mechanism, is proposed as a measure to quantify non-normality. When the operator is normal but not self-adjoint, a phase difference between the forcing and response modes is indicated. In flows where there is downstream spatial dependence of the base/mean, mean flow advection separates the spatial support of forcing and response modes which impacts the inner product and leads to non-normal amplification. Success of mean flow (linear) stability analysis for a particular frequency depends on the normality of amplification. If the amplification is normal, the resolvent operator can be rewritten in its dyadic representation to reveal that the adjoint and forward stability modes are proportional to the forcing and response resolvent modes at that frequency. If the amplification is non-normal, then resolvent analysis is required to understand the origin of observed flow structures. Eigenspectra and pseudospectra are used to characterize these phenomena. Two test cases are studied: low Reynolds number cylinder flow and turbulent channel flow. The first deals mainly with normal mechanisms hence the success of both classical and mean stability analysis with respect to predicting the critical Reynolds number and global frequency of the saturated flow. Quantification of non-normality using the inverse inner product of the leading forcing and response modes agrees well with the product of the resolvent norm and distance between the imaginary axis and least stable eigenvalue. An approximate wavemaker computed from the resolvent modes scales with the length of the mean recirculation bubble. In the case of turbulent channel flow, structures result from both normal and non-normal mechanisms, suggesting that both are necessary elements to sustain turbulence. Mean shear is exploited most efficiently by stationary disturbances while bounds on the pseudospectra illustrate how non-normality is responsible for the most amplified disturbances at spatial wavenumbers and temporal frequencies corresponding to well-known turbulent structures. Some implications for flow control are discussed.

%\begin{description}
%\end{description}
\end{abstract}

\maketitle

%\tableofcontents

\section{Introduction}
\label{sec:intro}

\subsection{Background}
Decomposing unsteady and turbulent flows into low-rank models is an important step towards the realization of closed-loop flow control, which has proven to be elusive (at least in part) due to the high degrees of freedom in most flow systems. Despite the inherent complexity, a time-averaged flow, or mean, that is statistically stationary can often be defined and leveraged using the eigenvalue spectrum of the governing Navier-Stokes equations (NSE) to educe the frequencies, i.e. the imaginary part of the eigenvalues, and shapes of coherent structures which appear in the flow. Recent studies have demonstrated the success of mean flow stability analysis for a variety of flows including thermosolutal convection~\cite{Turton15}, turbulent jets~\cite{Gudmundsson11, Oberleithner14, Schmidt17}, and flow over a backward facing step~\cite{Beneddine16}. There is also a significant body of work discussing stability analysis of the mean cylinder wake which was shown by \citet{Barkley06} to correctly identify the frequency of the globally unstable flow above the critical Reynolds number of $Re = 47$~\citep{Provansal87, Sreenivasan87, Noack94}. Notably, classical linear stability analysis of the base flow, which is an equilibrium solution of the NSE, at supercritical Reynolds numbers does not predict the correct observed frequency. The base (laminar) and mean (time-average of the fluctuating velocity field) profiles are differentiated because of the importance of nonlinearity in sustaining the latter.

Recent work has endeavored to explain why and when mean stability analysis is valid.~\citet{Barkley06} suggested that success corresponds to cases where the Reynolds stresses are unperturbed at order $\epsilon$ when considering infinitesimal perturbations $\epsilon \tilde{\boldsymbol{u}}(x,y)\text{exp}(\lambda t)$ to the mean flow solution. This was confirmed by \citet{Sipp07} who performed a weakly nonlinear expansion of the cylinder base flow near criticality and determined that the nonlinear interaction of the leading global mode with its conjugate, i.e. the contribution to the mean Reynolds stresses, significantly outweighed the interaction of the mode with itself leading to additional harmonics of the wavenumber. \citet{MLugo14} provided a self-consistent framework for the cylinder mean flow by approximating the divergence of the Reynolds stresses with the leading global mode and its conjugate.

~\citet{Sipp07} used open cavity flow as a counter example to the validity of mean stability analysis where the predicted frequencies do not match direct numerical simulation (DNS) of the flow. This discrepancy can be attributed to the non-normality of the flow which leads to non-orthogonality of the global modes and sensitivity of the spectrum to perturbation of the operator~\cite{Trefethen93}. In such circumstances, the behavior of the system can be more accurately characterized by consideration of the properties of the pseudospectrum of the LNS operator using resolvent analysis~\cite[e.g.][]{Trefethen93,Schmid01} rather than the spectrum itself. \citet{Jovanovic05} formulated the linearized problem for laminar channel flow in input-output terms, where the resolvent operator constitutes the transfer function between them, considering the component-wise transfer from harmonic exogenous disturbance or forcing (input) to velocity response (output). There is also a broad literature considering stochastic forcing, \cite[e.g.][]{Farrell93}, and the initial condition, transient growth problem, \cite[e.g][]{Butler92}.

\citet{McKeon10} and \citet{Hwang10} considered the resolvent reformulated with respect to the turbulent mean flow for canonical turbulent wall flows. The latter authors employed an eddy viscosity to account for the action of the Reynolds stresses, while the former analysis extends the approach to include the nonlinear terms as the input forcing to the linear operator, i.e. closing the feedback loop. \citet{McKeon10} performed a singular value decomposition of the resolvent to identify the inputs giving rise to the most amplified responses which are ranked by their gain (singular value). The approach has been extended to non-parallel flows ~\cite[e.g.][]{Lu14, Beneddine16, Jeun16, Schmidt17b}. \citet{Beneddine16} concluded that mean stability analysis was valid when the dominant singular value of the resolvent operator was significantly greater than the others at a given frequency and that this condition holds for flows where there is a dominant convective instability mechanism and an eigenvalue which is nearly marginally stable. In such circumstances, it was shown that the eigenmodes are proportional to the resolvent response modes.

\subsection{Motivation and scope of the study}

A broad review of the approaches discussed above can be found in \citet{Taira17}. Given the lack of a comprehensive understanding of when spectral and pseudospectral analyses lead to perturbations with characteristics that are observed in real flows, and when the base or mean flow will lead to success, our approach is to synthesize the characteristics of the LNS operator with a view to answering specific questions.

One of the main objectives of this paper is to understand the conditions under which eigenanalyses and resolvent analyses yield the same results, i.e. when spectral and pseudospectral analyses reveal the same most amplified perturbations. Furthermore, we seek to address when each method is likely to lead to successful prediction of flow features, and whether the properties of the base or mean flow (which are inputs to form the linear operators for both eigen- and resolvent analyses) indicate when these analyses coincide. This leads to deeper questions about the resolvent operator and the types of amplification mechanisms it admits.

It is well known that the response of a system to harmonic input (forcing) can occur due to normal and/or non-normal mechanisms. The former corresponds to excitation in the vicinity of an eigenvalue (with exponential growth if the eigenvalue is unstable) while the latter is due to nonmodal effects associated with the sensitivity of the spectrum to perturbation. Subsequent to the work of \citet{Jovanovic05}, \citet{Marquet09} and~\citet{Brandt11} investigated the distribution of energy and relative phase between velocity components in analyses about laminar, base flows in recirculation bubbles and the flat plate boundary layer, respectively. These studies distinguished between component-type non-normalities which distribute energy in different velocity components and convective non-normalities, which separate the spatial support of forcing upstream of the response. The roots of these non-normalities are the mean shear and mean flow advection terms, respectively, in the linearized NSE. These terms also result in the Orr mechanism~\cite{Orr07} which reorients upstream-leaning forcing modes with the mean shear such that the response modes are leaning downstream~\cite{Farrell87}. In this paper we extend the analysis to include turbulent mean flows and quantify non-normality based on inner products between the resolvent forcing and response modes. The special case of limited spatial overlap of the forcing and response modes, which is known as a wavemaker~\cite[e.g.][]{Pier01} and arises when there is sufficiently strong reverse flow~\cite{Huerre85}, is also addressed.

Non-normality can be investigated through the lens of the pseudospectrum of the LNS operator. Most previous studies have only investigated the pseudospectra of parallel base flows~\citep[e.g.][]{Trefethen93, Reddy93, Trefethen99, Schmid01, Schmid07, Schmid14}. In this study, pseudospectral analysis is extended to mean flows with or without spatial development. The contribution from normality of the operator, or resonance, can be quantified from the spectrum alone by computing the inverse distance between the imaginary axis and the nearest eigenvalue. The contribution from non-normality of the operator can be quantified by bounds of the pseudospectrum which may be dramatically different from the spectrum if the operator is non-normal and thus sensitive to perturbations. For complex flows where a variety of frequencies and wavenumbers are active, both normal and non-normal mechanisms may be simultaneously present and dynamically important.

Lastly, we consider the conditions under which the resolvent operator can be considered to be low-rank. Here we use the term low-rank to describe an operator whose behavior can be well approximated by a low-rank operator as opposed to the meaning that the operator is of low numerical rank. \citet{McKeon10} identified low-rank characteristics of the resolvent for turbulent wall flows via the rapid decrease of singular value with increasing order of the singular value decomposition. They attributed this to the dominance of specific physical mechanisms that are efficiently captured through representation in terms of spatio-temporal modes. Subsequent work~\citep[e.g.][]{Gomez16, Beneddine16} has exploited the low-rank approximation of the resolvent to construct reduced-order models of complex flows. We show that the question of rank of the resolvent at a particular frequency can be investigated by analyzing the pseudospectrum as well as the ratio of the largest to smallest terms in the resolvent operator itself.

\subsection{Choice of flows and outline of the paper}

We consider low Reynolds number cylinder flow and turbulent channel flow to demonstrate the contributions of this paper. Flow around a circular cylinder is an example of an oscillator flow~\citep{Huerre98} which has intrinsic dynamics that are insensitive to background noise. The flow exhibits a region of absolute instability~\cite{Huerre85, Monkewitz88} which can be approximated using the wavemaker~\citep[e.g.][]{Giannetti07}. Using resolvent analysis and plotting bounds of the pseudospectrum for a given perturbation magnitude, it is evident that resonance accounts for the bulk of the amplification. The impact of mean flow advection, however, which separates the spatial support of the forcing and response modes, introduces convective non-normality and leads to a resolvent norm which is appreciably larger than the contribution from resonance alone. In the case of turbulent channel flow, which is a noise-amplifier~\cite{Huerre98}, the choice of temporal frequency and spatial wavenumbers has an impact on the influence of non-normality. We analyze several structures which are either highly amplified or representative of known turbulent structures in order to establish the role of normal and non-normal mechanisms as well as generalize when low-rank behavior can be expected.

The rest of the paper is organized as follows. In Section~\ref{sec:equations}, the governing equations are derived for the linear operators which form the basis of stability and resolvent analysis. In Section~\ref{sec:amp}, the resolvent operator is rewritten to differentiate normal and non-normal amplification mechanisms. A model LNS operator and its resolvent without spatial dependence is used to isolate the effects of individual terms in the NSE on the forcing and response modes. The effect of spatial dependence is discussed particularly under the influence of mean shear and reverse flow. Finally, the resolvent operator is rewritten using its dyadic representation to formally relate stability and resolvent modes and show when the rank-1 approximation is appropriate for normal mechanisms. Application of the findings to circular cylinder flow is considered in Section~\ref{sec:cylinder}; the numerical methods are outlined and resolvent analysis is applied to both base and mean flows. The results are compared to those from a stability analysis. Section~\ref{sec:turbulence} studies the observations from earlier sections in the context of turbulent channel flow. The influence of the wall-normal height and spatial and temporal wavenumbers are shown to play a major factor in the type of amplification mechanisms which dominate as well as whether or not the resolvent is low-rank. Conclusions are presented in Section~\ref{sec:conclusion} along with the implications for reduced-order modeling and control.

\section{Governing Equations}
\label{sec:equations}

The relevant operators for the analyses that follow are derived from the incompressible NSE which are non-dimensionalized by the characteristic length and velocity scales, $L$ and $U$:
\begin{subequations}
\begin{align}
\partial_t \boldsymbol{u} + \boldsymbol{u} \cdot \nabla \boldsymbol{u} &= -\nabla p + Re^{-1} \nabla^2
\boldsymbol{u} \label{eq:NS momentum}  \\  \nabla \cdot \boldsymbol{u} &= 0.
\end{align}
\label{eq:NS}
\end{subequations}
The states $\boldsymbol{u}(\boldsymbol{x},t)$ and $p(\boldsymbol{x},t)$ are the spatially- and temporally-varying velocity and pressure fields, respectively (explicit statement of the spatial and temporal dependences will be dropped hereon for conciseness), while $Re$ is the Reynolds number based on $L$ and $U$. After Reynolds-decomposing the states into a stationary temporal mean (denoted by an overline) and a fluctuating component (denoted by a prime), one obtains the mean flow equations:
\begin{subequations}
\begin{align}
\overline{\boldsymbol{u}} \cdot \nabla \overline{\boldsymbol{u}} + \nabla \overline{p} - Re^{-1} \nabla^2 \overline{\boldsymbol{u}} &= - \overline{\boldsymbol{u}' \cdot \nabla \boldsymbol{u}'} \label{eq:mean momentum} \\
\nabla \cdot \overline{\boldsymbol{u}} &= 0.
\end{align} \label{eq:mean}
\end{subequations}
For an exact solution of the NSE, i.e. a true \textit{base flow}, the divergence of the Reynolds stress tensor, or the right-hand term in Equation~\ref{eq:mean momentum}, is zero. Henceforth we identify the temporal mean in such a case by $\overline{\boldsymbol{u}} = \boldsymbol{U}_0$. If the term is nonzero, however, $\overline{\boldsymbol{u}}$ does not constitute an exact solution and the action of Reynolds stresses must be taken into account to satisfy the NSE. We will refer to $\overline{\boldsymbol{u}}$ in this case as a (turbulent) \textit{mean flow}.

Subtracting the mean momentum equations (Equation~\ref{eq:mean}) from the NSE (Equation~\ref{eq:NS}) yields the following for the fluctuating quantities:
\begin{subequations}
\begin{align}
\partial_t \boldsymbol{u}' +  \overline{\boldsymbol{u}} \cdot \nabla \boldsymbol{u}' + \boldsymbol{u}' \cdot \nabla \overline{\boldsymbol{u}} + \nabla p' -Re^{-1} \nabla^2 \boldsymbol{u}' =  -\boldsymbol{u}' \cdot \nabla\boldsymbol{u}' + \overline{\boldsymbol{u}' \cdot \nabla \boldsymbol{u}'} = \boldsymbol{f}' \label{eq:fluctuating momentum} \\
\nabla \cdot \boldsymbol{u}' = 0.
\label{eq:fluctuating}
\end{align}
\end{subequations}
Equation~\ref{eq:fluctuating momentum} has been written such that all linear terms appear on the left-hand side. The nonlinear terms on the right-hand side can be lumped together as a forcing $\boldsymbol{f}'$ without loss of generality. The literature devoted to analyzing these linearized equations is broad and covers a range of flows. We recap here the analyses required for the subsequent development and refer the reader to recent reviews and contributions~\citep[][]{Chomaz05, Schmid07, Sipp10, Schmid14, McKeon17} for further information.

Our development considers global and spatially periodic modes in the context of bluff body and wall-bounded turbulent flows, respectively. The conclusions, nevertheless, are applicable when using the parabolized stability equations (PSE), \cite[e.g.][]{Beneddine16}, although PSE is appropriate only for the treatment of convective instability and not absolute instability \citep{Herbert97}. They are also applicable to both base and mean flows although there are important differences between the results for these profiles which will be drawn out. The temporal mean velocity profiles are known throughout the domain from numerical simulation for the cylinder flow while they are obtained via an eddy viscosity model~\citep[e.g.][]{Reynolds67} for the turbulent channel flow. We begin by considering the general case in which there exists invariance only in time. In other words, the analysis is performed in the frequency domain, such that mode shapes may be functions of all three spatial dimensions. Decomposition into spatial wavenumbers is applied in the case of wall-bounded turbulent flows with streamwise and spanwise periodicity in Section~\ref{sec:turbulence}.

\subsection{Eigenmode decomposition}

The classical (temporal) linear stability analysis, detailed, for example, in \cite{Schmid01}, proceeds under the assumption of small perturbations to the steady state. It has been performed relative to both base and mean flows in order to determine frequencies subject to exponential growth (instability). See e.g. \cite{Taira17} for further details.  Substituting perturbations of the form
\begin{equation}\label{eq:eigperts}
\boldsymbol{u}(\boldsymbol{x},t) = \overline{\boldsymbol{u}}(\boldsymbol{x}) +\epsilon \tilde{\boldsymbol{u}}(\boldsymbol{x}) \exp(\lambda t),
\end{equation}
where $\epsilon \ll 1$, into Equation~\ref{eq:NS} yields at $\mathcal{O}(\epsilon)$:
\begin{subequations}
\begin{align}
\lambda \tilde{\boldsymbol{u}} & = -\overline{\boldsymbol{u}} \cdot \nabla \tilde{\boldsymbol{u}} - \tilde{\boldsymbol{u}} \cdot \nabla \overline{\boldsymbol{u}} - \nabla \tilde{p} + Re^{-1} \nabla^2 \tilde{\boldsymbol{u}}
\label{eq:stability momentum} \\
\nabla \cdot \tilde{\boldsymbol{u}} & = 0,
\end{align} \label{eq:stability}
\end{subequations}
where a tilde will be used to denote that a stability analysis has been performed. Written explicitly for base flows in operator form, one obtains
\begin{equation}
\lambda \boldsymbol{B} \left( \begin{array}{c} \tilde{\boldsymbol{u}} \\ \tilde{p} \end{array} \right) = \boldsymbol{A} \left( \begin{array}{c} \tilde{\boldsymbol{u}} \\ \tilde{p} \end{array} \right),
\end{equation}
where $\boldsymbol{A}$ is the LNS operator with respect to the base flow,
\begin{equation}
 \boldsymbol{A} = \left(\begin{array}{cc} -\boldsymbol{U}_0 \cdot \nabla() - () \cdot \nabla \boldsymbol{U}_0 + Re^{-1} \nabla^2()& -\nabla() \\ \nabla \cdot () & 0 \end{array} \right),
\label{eq:base operator}
\end{equation}
and
\begin{equation}
 \boldsymbol{B} = \left( \begin{array}{cc} 1 & 0 \\ 0 & 0 \end{array}  \right).
\end{equation}
Similarly, we use $\boldsymbol{L}$ to denote the LNS operator with respect to the mean flow, such that for mean flow stability analysis
\begin{equation}
\lambda \boldsymbol{B} \left( \begin{array}{c} \tilde{\boldsymbol{u}} \\ \tilde{p} \end{array} \right) = \boldsymbol{L} \left( \begin{array}{c} \tilde{\boldsymbol{u}} \\ \tilde{p} \end{array} \right),
\end{equation}
with
\begin{equation}
\boldsymbol{L} = \left(\begin{array}{cc} -\overline{\boldsymbol{u}} \cdot \nabla() - () \cdot \nabla \overline{\boldsymbol{u}} + Re^{-1} \nabla^2() & -\nabla() \\ \nabla \cdot () & 0 \end{array} \right).
\label{eq:mean operator}
\end{equation}
The resulting eigenvalue problems for base and mean flows are summarized in the upper row of Table~\ref{tab:operator equations}. Stability analysis is based on the spectrum of the LNS operator and $\lambda$ is an indicator of the linear stability of a given profile. The eigenvectors may be used as a basis for modal decomposition.

\begin{table}
\begin{center}
\begin{tabular}{ r | c | c | }
\multicolumn{1}{r}{}
 &  \multicolumn{1}{c}{Base Flow}
 & \multicolumn{1}{c}{Mean Flow} \\
\cline{2-3}
& & \\
Stability Analysis & $ \lambda \boldsymbol{B} \left( \begin{array}{c} \tilde{\boldsymbol{u}} \\ \tilde{p} \end{array} \right) = \boldsymbol{A} \left( \begin{array}{c} \tilde{\boldsymbol{u}} \\ \tilde{p} \end{array} \right)$ & $ \lambda \boldsymbol{B} \left( \begin{array}{c} \tilde{\boldsymbol{u}} \\ \tilde{p} \end{array} \right) = \boldsymbol{L} \left( \begin{array}{c} \tilde{\boldsymbol{u}} \\ \tilde{p} \end{array} \right) $ \\
& & \\
\cline{2-3}
& & \\
Resolvent Analysis & ~$i\omega \boldsymbol{B}\left( \begin{array}{c} \hat{\boldsymbol{u}} \\ \hat{p} \end{array} \right) = \boldsymbol{A} \left( \begin{array}{c} \hat{\boldsymbol{u}} \\ \hat{p} \end{array} \right) + \boldsymbol{C}\hat{\boldsymbol{f}}$~ & ~$i\omega \boldsymbol{B} \left( \begin{array}{c} \hat{\boldsymbol{u}} \\ \hat{p} \end{array} \right) = \boldsymbol{L} \left( \begin{array}{c} \hat{\boldsymbol{u}} \\ \hat{p} \end{array} \right) + \boldsymbol{C}\hat{\boldsymbol{f}}$~ \\
& & \\
\cline{2-3}
\end{tabular}
\end{center}
\caption{Operator form of the equations for stability and resolvent analyses. Variables with a tilde correspond to stability analysis while a caret indicates resolvent analysis.} \label{tab:operator equations}
\end{table}

The adjoint NSE have been derived by, e.g.~\cite{Luchini14}, and the linearized operators can be written for both base and mean flows, $\boldsymbol{A}^{*}$ and $\boldsymbol{L}^{*}$, respectively, for the adjoint variables $\tilde{\boldsymbol{v}}$ and $\tilde{q}$,
\begin{eqnarray}
\boldsymbol{A}^{*} = \left( \begin{array}{cc} \boldsymbol{U}_0 \cdot \nabla () - () \cdot (\nabla \boldsymbol{U}_0)^{*} + Re^{-1}\nabla^2 () & \nabla () \\ \nabla \cdot () & 0 \end{array} \right),
\label{eq:adjoint operator_base}\\
\boldsymbol{L}^{*} = \left( \begin{array}{cc} \overline{\boldsymbol{u}} \cdot \nabla () - () \cdot (\nabla \overline{\boldsymbol{u}})^{*} + Re^{-1}\nabla^2 () & \nabla () \\ \nabla \cdot () & 0 \end{array} \right).
\label{eq:adjoint operator_mean}
\end{eqnarray}
The operators satisfy
\begin{eqnarray}
\left<\tilde{\boldsymbol{u}},\boldsymbol{A}\tilde{\boldsymbol{v}}\right> = \left<\boldsymbol{A}^*\tilde{\boldsymbol{u}},\tilde{\boldsymbol{v}}\right>,\\
\left<\tilde{\boldsymbol{u}},\boldsymbol{L}\tilde{\boldsymbol{v}}\right> = \left<\boldsymbol{L}^*\tilde{\boldsymbol{u}},\tilde{\boldsymbol{v}}\right>,
\end{eqnarray}
where $<,>$ is the scalar product associated with the energy in the whole domain.

For a general operator, $\boldsymbol{T}$, that is normal, i.e. $\boldsymbol{T}\boldsymbol{T}^{*} = \boldsymbol{T}^{*}\boldsymbol{T}$, the eigenvectors of $\boldsymbol{T}$ corresponding to distinct eigenvalues are orthogonal although the eigenvalues may be complex. Self-adjoint operators $(\boldsymbol{T} = \boldsymbol{T}^{*})$, on the other hand, have orthogonal eigenvectors and real eigenvalues. In general, the LNS operators $\boldsymbol{A}$ and $\boldsymbol{L}$ are neither self-adjoint nor normal. These phenomena account for the differences between Equations~\ref{eq:base operator},~\ref{eq:mean operator} and Equations~\ref{eq:adjoint operator_base},~\ref{eq:adjoint operator_mean} and their influence is discussed in Section~\ref{sec:amp}.

\subsection{Resolvent analysis}

For the more general case when the perturbation cannot be considered to be infinitesimal and the nonlinearity $\boldsymbol{f}'$ is retained, Equation~\ref{eq:fluctuating momentum} can be rewritten in terms of a transfer function between the forcing (input) and response state (output) \citep[e.g.][]{Schmid01,Jovanovic05,McKeon10}.  This transfer function is the (linear) resolvent operator, which can be defined around either the base or mean flow (see Table~\ref{tab:operator equations}). Endogenous nonlinear terms in Equation~\ref{eq:fluctuating} when $\epsilon$ is not required to be small or exogenous forcing can be treated equally well via $\boldsymbol{f}'$, although the interpretation of the resulting system is different.

For harmonic forcing and response at temporal frequency $\omega$,
\begin{equation}
\boldsymbol{f}' = \hat{\boldsymbol{f}}~\text{exp}(i\omega t), \quad \boldsymbol{u}' = \hat{\boldsymbol{u}}~\text{exp}(i\omega t)
\end{equation}
and a base flow profile, $\overline{\boldsymbol{u}} = \boldsymbol{U}_0$,
\begin{equation}
\hat{\boldsymbol{u}} = \mathcal{H}(\omega) \hat{\boldsymbol{f}},
\label{eq:resolvent operator}
\end{equation}
where the caret denotes that the perturbation is associated with a resolvent analysis. The resolvent operator is given by
\begin{equation}
\mathcal{H}(\omega) = \boldsymbol{C}^T(i\omega \boldsymbol{B} - \boldsymbol{A})^{-1}\boldsymbol{C},
\end{equation}
where
\begin{equation}
\boldsymbol{C} = \left(\begin{array}{c} 1 \\ 0 \end{array} \right).
\end{equation}
Similarly, for a mean flow,
\begin{equation}
\mathcal{H}(\omega) = \boldsymbol{C}^T(i\omega \boldsymbol{B} - \boldsymbol{L})^{-1}\boldsymbol{C}.
\end{equation}
It should be noted that the sense of the imaginary and real parts of $\omega$ is reversed in the resolvent analysis relative to the definition customary to the stability literature of Equation~\ref{eq:eigperts}: here the real part of $\omega$ is the frequency associated with a mode while the imaginary part is set to zero as only neutral disturbances are considered. In stability analysis, the imaginary part of $\lambda$ is the frequency of the disturbance and the real part is the growth rate.

$\mathcal{H}(\omega)$ can be decomposed via a singular value decomposition (SVD), e.g. \cite{McKeon10}:
\begin{equation}
\mathcal{H}(\omega) = \boldsymbol{\Psi}(\omega)\boldsymbol{\Sigma}(\omega)\boldsymbol{\Phi}^{*}(\omega),
\label{eq:SVD}
\end{equation}
where $\boldsymbol{\Psi}$ and $\boldsymbol{\Phi}$ are the left and right singular vectors corresponding to the response and forcing modes, often called resolvent modes~\citep[see][]{McKeon10}, respectively. Both sets of singular vectors are guaranteed to be orthonormal bases and are ranked according to their gain, or singular value, contained in the diagonal matrix $\boldsymbol{\Sigma}$. The resolvent operator can thus be written as the sum of outer products of the left and right singular vectors

\begin{equation}
\mathcal{H}(\omega) = \sum_{j=1}^{\infty} \hat{\boldsymbol{\psi}}_j(\omega)\sigma_j(\omega)\hat{\boldsymbol{\phi}}^*_j(\omega).
\label{eq:resolvent full rank}
\end{equation}
$\mathcal{H}(\omega)$ is (approximately) low rank if
\begin{equation}
\sum_{j = 1}^p \sigma^2_j \approx \sum_{j = 1}^{\infty} \sigma^2_j,
\label{eq:low rank}
\end{equation}
where $\sigma_p \gg \sigma_{p+1}$ and $p$ is small \cite{Moarref13, McKeon17} . If the leading singular value is significantly greater than all others ($\sigma_1 \gg \sigma_2$) then the rank-1 approximation can be invoked and the resolvent is approximated by the outer product of the leading optimal response and forcing modes:
\begin{equation}
\mathcal{H}(\omega) \approx \sigma_1\hat{\boldsymbol{\psi}}_1 \hat{\boldsymbol{\phi}}_1^{*}.
\label{eq:rank 1}
\end{equation}
The physical interpretation of the resolvent response modes is the response to forcing that results in a neutrally stable response, i.e. with the real component of frequency equal to zero. The singular value gives the input-output gain, here associated with the energy norm.

\subsection{Spectrum and pseudospectrum of the LNS operator}

Analyzing the resolvent corresponds to considering the spectrum of the perturbed LNS operator:
\begin{equation}
\Lambda_{\epsilon}(\boldsymbol{A}) = \{z \in \mathbb{C} : z \in \Lambda(\boldsymbol{A}+\boldsymbol{E}) ~\text{where}~\| \boldsymbol{E} \| \leq \epsilon \},
\label{eq:epsilon LNS}
\end{equation}
where $\Lambda_\epsilon$ is the pseudospectrum of $\boldsymbol{A}$ under a perturbation magnitude $\epsilon >0$ and $\| \cdot \| $ is the (operator) 2-norm \cite{Trefethen93, Reddy93, Taira17}. An equivalent definition is given by
\begin{equation}
\Lambda_{\epsilon}(\boldsymbol{A}) = \left\{z \in \mathbb{C} : \|(z \boldsymbol{I}-\boldsymbol{A})^{-1} \| \geq \epsilon^{-1} \right\} ~  \cup ~ \Lambda(\boldsymbol{A}),
\end{equation}
where $\Lambda = \Lambda_0 $ is the spectrum of $\boldsymbol{A}$. Throughout the paper, we will use $\Lambda$ to represent the set of eigenvalues and $\boldsymbol{\Lambda}$ the diagonal matrix of eigenvalues. If $\boldsymbol{A}$ is normal, $\Lambda_\epsilon$ can be interpreted as the set of points away from $\Lambda$ by only less than or equal to $\epsilon$ on the complex plane~\citep[see][]{Taira17}. If $\boldsymbol{A}$ is non-normal, this distance may be greater than $\epsilon$ signifying that an eigenvalue is sensitive to perturbation of the LNS operator.

For a given $z$, the resolvent norm $\|(z \boldsymbol{I}-\boldsymbol{A})^{-1} \|$ is equal to the largest value of $\epsilon^{-1}$ such that  $z$ is contained within $\Lambda_\epsilon$. The resolvent norm is (by definition) the maximum singular value of the resolvent operator, and quantifies the system's sensitivity to temporal forcing.
The neutrally stable response of the system to harmonic forcing is characterized by the value of the resolvent norm along the imaginary axis. For a stable, normal operator, the largest response occurs at the frequency corresponding to the imaginary part of the least stable eigenvalue $\lambda_{\text{ls}}$ and the resolvent norm is $1/\text{Real}(\lambda_{\text{ls}})$ since $\text{Real}(\lambda_{\text{ls}})$ is the minimum distance between the eigenvalue and the imaginary axis~\citep[][]{Chomaz05}. If the operator is marginally stable $(\text{Real}(\lambda_{\text{ls}}) \approx 0 )$, then $\epsilon \approx 0$ and the response is dominated by the corresponding eigenmode. For a stable, non-normal operator, the frequency and gain of the largest response is less predictable since it is necessary to find the smallest value of $\epsilon$ for which the pseudospectrum crosses the imaginary axis.

Having reviewed the derivations and interpretations of the two analyses that will be considered here, we now elaborate on the mathematical and physical origins of the associated amplification mechanisms and componentwise transfer. Simple example operators are used to illustrate the concepts, in the spirit of \citet{Gebhardt94}, before we consider the complete LNS operator.

\section{Mechanisms for amplification}
\label{sec:amp}
\subsection{Resonance and pseudoresonance}

The origin of the amplification mechanisms characterized by the resolvent norm can be identified by expanding the resolvent through an eigenvalue decomposition of the LNS operator,
\begin{equation}
\mathcal{H}(\omega) = \boldsymbol{C}^T(i\omega \boldsymbol{B} - \boldsymbol{V} \boldsymbol{\Lambda B} \boldsymbol{V}^{-1})^{-1} \boldsymbol{C}.
\label{eq:resolvent eigenvalue}
\end{equation}
Here $\boldsymbol{V}$ represents the matrix of eigenvectors of the LNS operator for either the base or mean flow profile and $\boldsymbol{\Lambda}$ the diagonal matrix of eigenvalues. These can be used to find an upper and lower bound for the resolvent norm~\cite[see][]{Schmid01}:
\begin{equation}
\| i\omega \boldsymbol{I} - \boldsymbol{\Lambda}\| ^{-1} \leq \| \mathcal{H}(\omega) \|  \leq  \underbrace{\|\boldsymbol{V} \| \| \boldsymbol{V}^{-1}\|}_{\text{pseudoresonance}} \underbrace{\|i\omega \boldsymbol{I} - \boldsymbol{\Lambda} \|^{-1}}_\text{resonance}.
\label{eq:resolvent eigenvalue}
\end{equation}

Considering first the far-righthand term in Equation~\ref{eq:resolvent eigenvalue}, it is clear that forcing in the vicinity of an eigenvalue, i.e. $\omega=\lambda$, is likely to lead to amplification due to resonance. This is a normal phenomenon, predictable from an eigenanalysis. Large amplification also arises in the event of pseudoresonance when the condition number $\kappa = \| \boldsymbol{V} \| \|\boldsymbol{V}^{-1}\|$ is large due to non-orthogonality of the eigenvectors, a consequence of the non-normal nature of $\boldsymbol{L}$ and hence $\mathcal{H}(\omega)$. In the formulation of Equation~\ref{eq:resolvent eigenvalue}, the resolvent therefore contains both the amplification mechanisms associated with the eigenvalue spectrum accessed via eigenanalysis (normal mode linear stability analysis) and the non-normal amplification that is possible when the eigenvectors are not orthogonal to each other. In what follows, we will identify ``optimal pseudoresonance" as the condition under which the amplification in Equation~\ref{eq:resolvent eigenvalue} is associated with pseudoresonance, i.e. $\kappa \to \infty$, to be compared with the resonant case in which $\kappa = 1$.

\begin{figure}
       \centering
       \subfloat[]{\includegraphics[scale=0.27,trim=1cm 0cm 1cm 0cm, clip=true]{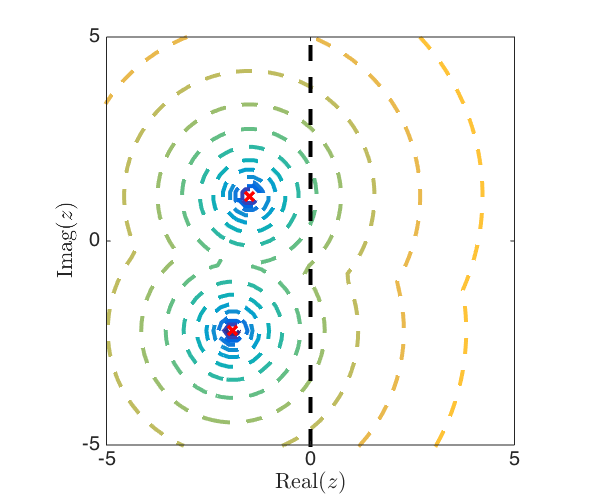}}
       \subfloat[]{\includegraphics[scale=0.27,trim=0.5cm 0cm 0cm 0cm, clip=true]{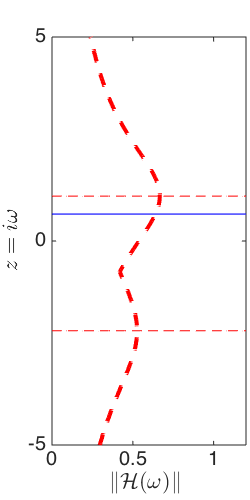}}
       \subfloat[]{\includegraphics[scale=0.27,trim=1cm 0cm 1cm 0cm, clip=true]{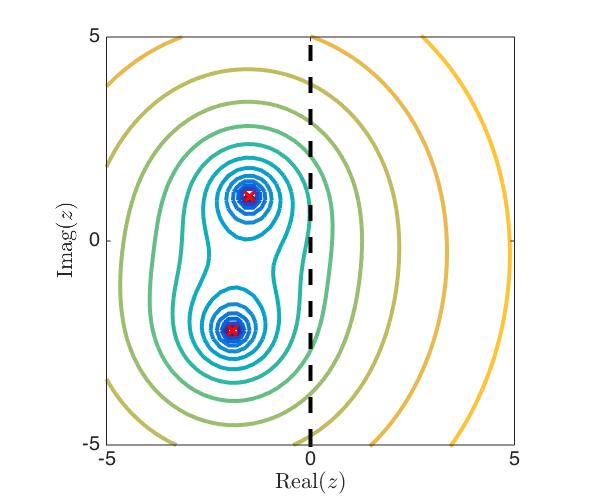}}
       \subfloat[]{\includegraphics[scale=0.27]{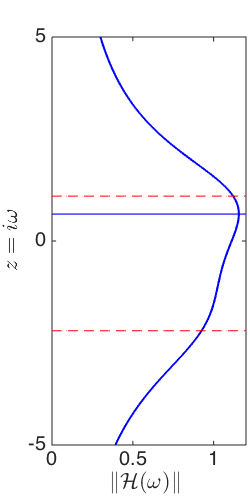}}
\caption{Comparison of the pseudospectra and resolvent norm for the operators given in Equations~\ref{eq:spectral_operator} and \ref{eq:pseudo operator}, which have the same eigenvalues. (a, b) normal operator $\boldsymbol{S}$, (c, d) non-normal operator $\boldsymbol{P}$. The eigenvalues, i.e. the eigenspectrum, are marked by red crosses and color contours outline the bounds of the perturbed spectrum for constant perturbation magnitudes in (a, c). The dashed contours in (a) reflect that the pseudospectra are circles centered on the eigenvalues, which is not true for the non-normal operator in (c). The resolvent norm in each case (b, d) reflects the value of these contours along the imaginary axis. Red, dashed horizontal lines indicate the resonant frequencies of the operator, i.e. the frequencies corresponding to the eigenvalues, while the blue, solid horizontal line represents the most highly amplified frequency in the non-normal case.}
\label{fig:pseudospectra}
\end{figure}

To demonstrate the amplification characteristics of the resolvent for normal and non-normal operators, we sketch in Figure~\ref{fig:pseudospectra} the pseudospectra for simple example operators,
\begin{eqnarray}
\boldsymbol{S} =  \left( \begin{array}{cc} -1.5 + 1.1i & 0 \\ 0 & -1.9-2.2i \end{array} \right),\label{eq:spectral_operator}\\
\boldsymbol{P} = \left( \begin{array}{cc} -1.5 + 1.1i & 5 \\ 0 & -1.9-2.2i \end{array} \right),
\label{eq:pseudo operator}
\end{eqnarray}
where $\boldsymbol{S}$ is a normal operator containing only the eigenvalues of the non-normal operator $\boldsymbol{P}$. Level curves of $\epsilon$ for operators $\boldsymbol{S}$ and $\boldsymbol{P}$ satisfy

\begin{equation}
\Lambda_{\epsilon} (\boldsymbol{S}) = \{z \in \mathbb{C}: \| (z \boldsymbol{I} - \boldsymbol{S} )^{-1} \| \geq \epsilon^{-1}\},
\end{equation}
and
\begin{equation}
\Lambda_{\epsilon} (\boldsymbol{P}) = \{z \in \mathbb{C}: \| (z \boldsymbol{I} - \boldsymbol{P} )^{-1} \| \geq \epsilon^{-1}\},
\end{equation}
respectively.

For a normal operator such as $\boldsymbol{S}$, $\kappa = 1$ and the level curves of $\epsilon$ are proportional to the distance from the closest eigenvalue. The resolvent norm for a particular $\omega$ is inversely proportional to the distance from $i \omega$ to the nearest eigenvalue. The spectrum and pseudospectra of $\boldsymbol{S}$ are shown in Figure~\ref{fig:pseudospectra}(a); there are two stable eigenvalues denoted by red crosses, and the pseudospectra consist of circular contours centered on the two eigenvalues. Since both eigenvalues are significantly damped, amplification due to resonance is not possible and the magnitude of the resolvent norm is less than one (Figure~\ref{fig:pseudospectra}(b)). Moreover, the eigenvalue and singular value decompositions of $\boldsymbol{S}$ yield parallel basis functions, and the singular values are simply the magnitude of the eigenvalues.

Operator $\boldsymbol{P}$, however, is non-normal due to the non-zero off-diagonal term and, with reference to Equation~\ref{eq:resolvent eigenvalue}, $\kappa > 1$. The shifts of the eigenvalues of the perturbed operator are not proportional to $\epsilon$, as indicated by the pseudospectrum isocontours in Figure~\ref{fig:pseudospectra}(c), and the resolvent norm of Figure~\ref{fig:pseudospectra}(d) is appreciably larger than that of the normal case in Figure~\ref{fig:pseudospectra}(a), with values exceeding one. Furthermore, the maximum value of the resolvent norm occurs at a non-resonant frequency, $\omega = 0.66$.  Amplification is possible for a linearly stable operator due to pseudoresonance under forcing at any frequency for which $\| \mathcal{H}(\omega) \| > 1$. It is important to add that even when the primary contribution to amplification is a normal mechanism, non-normality can still contribute to exacerbate the response. For example, the amplification at the frequencies of the eigenvalues in Figure~\ref{fig:pseudospectra} is higher for operator $\boldsymbol{P}$ than it is for the purely normal operator $\boldsymbol{S}$. That being the case, the right-hand side of Equation~\ref{eq:resolvent eigenvalue} may be large due to one or both terms in the product.

\subsection{Eigenvectors, singular functions, and non-normality of the operator}

We specialize now to operators with similar behavior to the LNS system and consider the associated features of eigenvector and singular value decompositions. We choose a model operator with characteristics similar to the LNS, in the vein of that explored by \citet{Gebhardt94} but without specifying numerical values. The LNS for the two-dimensional velocity field associated with a one-dimensional base or mean flow variation takes a similar form to $\boldsymbol{M}$, where
\begin{equation}
\boldsymbol{M} =  \left( \begin{array}{cc} m_1 & d \\ 0 & m_2 \end{array} \right).
\label{eq:L primitive}
\end{equation}

By selecting a one-dimensional operator, we have elected to neglect spatial (streamwise and spanwise) dependence of the base flow and therefore the modes themselves here. Nevertheless, the impact of the various types of term in the LNS operator on the resolvent modes can be modeled. Here $\text{Real}(m_j) < 0$ is analogous to the stabilizing role of viscosity through the $Re^{-1} \nabla^2 () $ term and $\text{Imag}(m_j)$ represents mean flow advection through the $-\overline{\boldsymbol{u}} \cdot \nabla ()$ term. $d$ is real and is analogous to mean shear $() \cdot \nabla \overline{\boldsymbol{u}}$, which here is equal to the gradient in the 2-direction of mean flow in the 1-direction. Thus in this simple 2-by-2 example, $d$ models the lift-up mechanism~\citep{Landahl80} by coupling forcing in the $n_2$-direction (second component of the vector) with a response in the $n_1$-direction (first component of the vector).

The resolvent of $\boldsymbol{M}$ is:
\begin{equation}\label{eq:resolvent_model}
\mathcal{H}(\omega) = \left( \begin{array}{cc} -1/(m_1-i\omega) & d/[(m_1-i\omega)(m_2-i\omega)] \\ 0 & -1/(m_2-i\omega) \end{array} \right).
\end{equation}
In order to isolate the effect of non-normality introduced via the off-diagonal term in Equation~\ref{eq:resolvent_model}, the eigenvalues are assumed to be real and we limit the immediate development to stationary disturbances ($\omega = 0$) to eliminate the remaining imaginary terms, such that
\begin{equation}\label{eq:resolvent_model}
\mathcal{H}(\omega = 0) = \left( \begin{array}{cc} -1/(m_1) & d/[(m_1 m_2)] \\ 0 & -1/(m_2) \end{array} \right).
\end{equation}

\begin{figure}[!htb]
        \centering
        \subfloat[]{\includegraphics[scale=0.33]{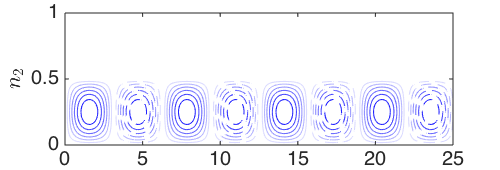}}
        \subfloat[]{\includegraphics[scale=0.33]{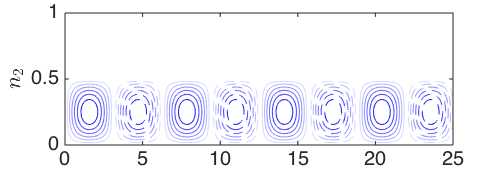}}

        \subfloat[]{\includegraphics[scale=0.33]{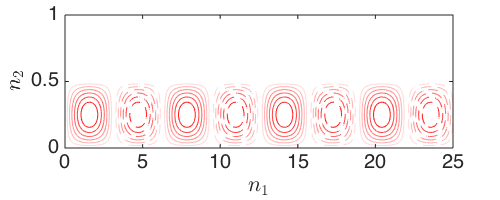}}
        \subfloat[]{\includegraphics[scale=0.33]{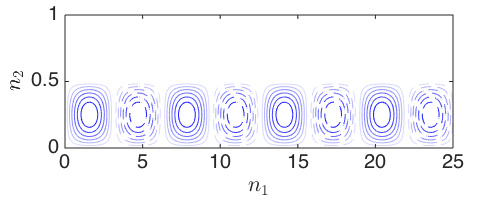}}
\caption{Cartoon of forcing and response modes corresponding to the operator given in Equation~\ref{eq:L primitive}, where the resolvent norm is large, for the limiting values of $d$. Panels (a, b) are the resonant forcing and response, respectively, for the case $d=0$ where the operator is self-adjoint with $\kappa = 1$ (and hence normal) while panels (c, d) are the optimal pseudoresonance forcing and response, respectively, for $d \to \infty$ and hence $\kappa \to \infty$. Positive/negative isocontours are denoted by solid/dotted lines and blue/red colors indicate streamwise/transverse components, in the $n_1, n_2$ directions, respectively. Each mode is nonzero in one velocity component only.}\label{fig:cartoon}
\end{figure}

For the limiting case of $d=0$, i.e. zero mean shear, $\boldsymbol{M}$ and its resolvent are self-adjoint and therefore normal. If the least stable eigenvalue has real part close to zero which would occur, say, if $m_1 \to 0$, the singular value decomposition can be simplified to
\begin{equation}
\lim_{m_1 \to 0} \text{SVD}(\mathcal{H}(\omega = 0)) = \left( \begin{array}{cc} 1 & 0  \\ 0 & 1  \end{array} \right) \left( \begin{array}{cc} \sigma_1 & 0 \\ 0 & \sigma_2 \end{array} \right) \left( \begin{array}{ccc} 1 & 0 \\ 0 & 1\end{array} \right),
\label{eq:a limit}
\end{equation}
where $\sigma_1/\sigma_2 \to \infty$, i.e. the resolvent is low-rank. Thus the response can be well predicted from the leading singular vectors $\hat{\boldsymbol{\psi}}_1$ and $\hat{\boldsymbol{\phi}}_1$. For a normal operator, these are identical to each other, $\hat{\boldsymbol{\psi}}_1 = \hat{\boldsymbol{\phi}}_1 = [1~0]^T$,  and identical to the corresponding eigenvectors. The inner product $| \hat{\boldsymbol{\phi}}_1^{*}\hat{\boldsymbol{\psi}}_1 |$ quantifies the componentwise correspondence between the forcing and response modes which, in this limit, is equal to unity. A schematic of the variation in the $n_1$ and $n_2$ directions of the forcing and response mode, i.e. the singular vectors, in this case is shown in Figure~\ref{fig:cartoon}(a-b).

The optimal pseudoresonance case occurs in the limit $d \to \infty$, for which:
\begin{equation}
\lim_{d \to \infty} \text{SVD}(\mathcal{H}(\omega = 0)) =  \left( \begin{array}{cc} 1-\gamma & -\delta \\ \delta & 1-\gamma \end{array} \right) \left( \begin{array}{cc} \sigma_1 & 0 \\ 0 & \sigma_2 \end{array} \right) \left( \begin{array}{cc} \delta & \gamma-1 \\ 1-\gamma & \delta \end{array} \right),
\label{eq:c limit svd}
\end{equation}
where $\sigma_1/\sigma_2 \to \infty$ and $\delta,~\gamma \to 0$. The constants $\delta$ and $\gamma$ are real and positive. The resolvent operator is still low-rank in this limit, but $\hat{\boldsymbol{\psi}}_1$ and $\hat{\boldsymbol{\phi}}_1$ are now orthogonal to each other and thus the inner product $| \hat{\boldsymbol{\phi}}_1^{*}\hat{\boldsymbol{\psi}}_1| \rightarrow 0$. The perturbation energy in the optimal forcing mode is concentrated in the second component of the vector while the perturbation energy in the optimal response mode is concentrated in the first component, as sketched in Figure~\ref{fig:cartoon}(c-d).

The analogous eigenvalue decomposition of (non-normal) $\boldsymbol{M}$ is
\begin{equation}
\lim_{d \to \infty} \text{EIG}(\boldsymbol{M}) =  \left( \begin{array}{cc} 1 & 1\\ 0 & \alpha \end{array} \right) \left( \begin{array}{cc} m_1 & 0 \\ 0 & m_2 \end{array} \right) \left( \begin{array}{cc} 1 & -1/\alpha \\ 0 & 1/\alpha \end{array} \right),
\label{eq:c limit eig}
\end{equation}
where $\alpha \to 0$ is a positive, real constant. Unlike the resolvent response modes which are orthogonal to one another, the eigenvectors are non-orthogonal, such that $\kappa \to \infty$. Thus, in this case the stability and resolvent modes are different.  Since the eigenvectors are nearly parallel, they both project equally well onto the optimal resolvent response mode. The same can be said for the projection of the adjoint modes, which are also nearly parallel, onto the optimal resolvent forcing mode. The optimal response and forcing modes, therefore, are linear combinations of multiple stability and adjoint modes, respectively. Furthermore, it is not clear from this decomposition that the resolvent operator is low-rank.

As outlined in Figure~\ref{fig:cartoon}, the model operator given in Equation~\ref{eq:L primitive} can be used to elucidate that the presence of mean shear, which introduces non-normality into the LNS operator, is to concentrate energy in different velocity components of the resolvent forcing and response modes. The model LNS operator does not need a base or mean flow with any special characteristics or spatial dependence other than an off-diagonal term to reveal this. Nevertheless, the same mechanism will be present in the case of more complex variation of shear, as is common in real flows. In the normal limit where there is no mean shear, the forcing and response are in the same velocity component. In the pseudoresonant limit, the forcing acts in the $n_2$ direction while the response is in the $n_1$ direction.

\subsection{Eigenvectors, singular functions, and self-adjointness of the operator}

In this section, we isolate the effect (or lack thereof) of self-adjointness on the characteristics of the resolvent modes. To this end, the mean shear term is set to zero, i.e. $d=0$ in Equation~\ref{eq:L primitive}, guaranteeing that the LNS and resolvent operators are normal:
\begin{eqnarray}
\boldsymbol{M} = \left( \begin{array}{cc} m_1 & 0 \\ 0 & m_2 \end{array} \right),
\label{eq:L convection} \\
\mathcal{H}(\omega) = \left( \begin{array}{cc} 1/(m_1- i \omega) & 0 \\ 0 & 1/(m_2 - i \omega) \end{array} \right).
\label{eq:H convection}
\end{eqnarray}
Again we will limit the discussion to stationary disturbances $(\omega = 0)$ as the presence of $i \omega$ in the resolvent operator will always guarantee that it is not self-adjoint.

If the eigenvalues are real, $\boldsymbol{M}$ and $\mathcal{H}(\omega = 0)$ are self-adjoint and the SVD of the resolvent is
\begin{equation}
\lim_{\text{Imag}(m_j)\to 0} \text{SVD}(\mathcal{H}(\omega = 0)) =  \left( \begin{array}{cc} 1 & 0 \\ 0 & 1 \end{array} \right) \left( \begin{array}{cc} m_1 & 0 \\ 0 & m_2 \end{array} \right) \left( \begin{array}{cc} 1 & 0 \\ 0 & 1 \end{array} \right).
\label{eq:self-adjoint svd}
\end{equation}
Equation~\ref{eq:self-adjoint svd} is similar to the normal limit case of Equation~\ref{eq:a limit} and illustrates that the resolvent forcing and response modes are identical.

\begin{figure}[!htb]
        \centering
        \subfloat[]{\includegraphics[scale=0.33]{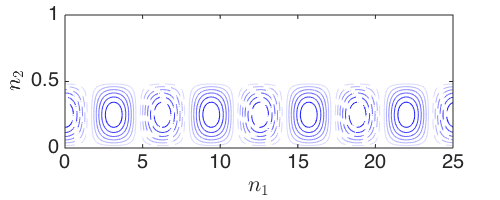}}
        \subfloat[]{\includegraphics[scale=0.33]{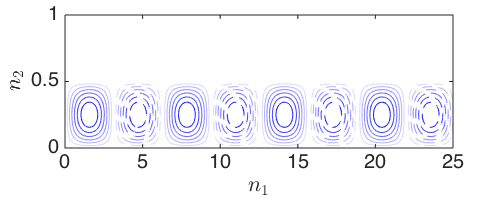}}
\caption{Cartoon illustrating the $\pi/2$ phase difference between (a) the resolvent forcing mode and (b) the resolvent response mode in Equation~\ref{eq:self-adjoint svd}. Positive/negative isocontours are denoted by solid/dotted lines, respectively.}\label{fig:cartoon_phase}
\end{figure}

The simplest way to demonstrate what happens when the resolvent is not self-adjoint is to assume that the eigenvalues are purely imaginary. In this case, the SVD of the resolvent is
\begin{equation}
\lim_{\text{Real}(m_j)\to 0} \text{SVD}(\mathcal{H}(\omega = 0)) =  \left( \begin{array}{cc} i & 0 \\ 0 & i \end{array} \right) \left( \begin{array}{cc} \| m_1 \| & 0 \\ 0 & \| m_2 \| \end{array} \right) \left( \begin{array}{cc} 1 & 0 \\ 0 & 1 \end{array} \right).
\label{eq:self-adjoint svd}
\end{equation}
Since the singular values of the resolvent are required to be real yet the eigenvalues are imaginary, the resolvent response modes must be $90^{\circ}$ out of phase from the resolvent forcing modes as is portrayed in Figure~\ref{fig:cartoon_phase}.

\subsection{Mean flow advection and the Orr mechanism}

In the case of an operator that is not self-adjoint and has non-zero mean shear, i.e. $d \ne 0$, the spatio-temporal manifestation of the Orr mechanism is observed. This mechanism results in disturbance amplification by the reorientation of an input tilted against the mean shear into a response which is aligned with the mean shear, with the maximum amplification in the transient case occurring when the disturbance is vertical.

\begin{figure}[!htb]
        \centering
        \subfloat[]{\includegraphics[scale=0.33]{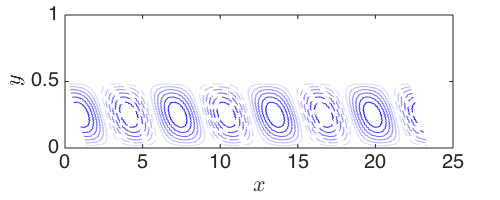}}
        \subfloat[]{\includegraphics[scale=0.33]{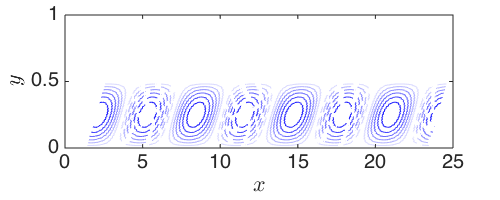}}
\caption{Cartoon illustrating the action of the Orr mechanism in resolvent modes. (a) The resolvent forcing mode leans upstream against the mean shear and (b) the resolvent response mode is rotated by the mean shear to lean downstream. Positive/negative isocontours are denoted by solid/dotted lines, respectively.}\label{fig:cartoon_Orr}
\end{figure}

In the continuously forced formulation of the present problem, the Orr mechanism is attributable to the space dependence of $\overline{\boldsymbol{u}}$ in the mean flow advection term. Assuming a parallel flow $(\overline{\boldsymbol{u}} = \overline{u}(y))$ with non-zero mean shear $(\partial \overline{u}/\partial y \neq 0)$ and traveling at the local mean velocity for a given $y_c$, the flow above $y_c$ moves towards the downstream direction while the flow below moves upstream in a relative sense. There is, therefore, a phase difference of $\pi$ across $y_c$ which manifests itself in the response modes~\citep[e.g.][]{McKeon17}. The decrease in phase results in the response modes leaning downstream, aligned with the mean shear. When considering the adjoint LNS operator $\boldsymbol{L}^{*}$, the mean flow advection acts in the upstream direction due to the sign change of $\boldsymbol{\overline{u}}\cdot \nabla ()$. The corresponding phase jump across $y_c$ is now in the opposite sense and this results in the forcing modes leaning upstream. Artificially removing mean flow advection from the resolvent operator suppresses the Orr mechanism. Figure~\ref{fig:cartoon_Orr} is a cartoon illustrating how the Orr mechanism results in the tilting of the resolvent forcing and response modes.

\subsection{Resolvent (approximate) wavemaker}\label{ssec:meanadvection}

For a spatially-varying base or mean flow, further statements can be made concerning the effect of the non-self-adjoint nature of the resolvent operator on the resolvent mode shapes, namely a difference in the spatial support of forcing and response modes, with the latter being downstream of the former. For a spatially-developing base or mean flow, normal mechanisms can be categorized in terms of either convective or absolute instability depending on the characteristics of the profile. The non-self-adjoint nature of $\boldsymbol{L}$ changes the influence of the convective terms in the adjoint operator, per Equations~\ref{eq:adjoint operator_base} and \ref{eq:adjoint operator_mean}, or, for the model operator of Equation~\ref{eq:L primitive},
\begin{equation}
\boldsymbol{M}^{*} = \left( \begin{array}{cc} m_1^* & 0\\ d & m_2^* \end{array} \right).
\label{eq:simple Astar}
\end{equation}
The direction of mean flow advection is reversed since the adjoint of the derivative operator introduces a negative sign implying that adjoint perturbations are transported upstream. Direct or forward perturbations, on the other hand, are transported downstream. For absolutely unstable flows, for which perturbations grow both upstream and downstream of the source, the advection term may no longer separate the spatial support of the forcing/adjoint and response/forward modes leading to regions of overlap at resonant frequencies. This region, known as the wavemaker, is traditionally computed from the eigenmodes, and is associated with non-zero values of $\mathcal{W}$, where
\begin{equation}
\mathcal{W}(\boldsymbol{x}_0) = \| \tilde{\boldsymbol{u}}(\boldsymbol{x}_0) \| \|\tilde{\boldsymbol{v}}(\boldsymbol{x}_0) \|,
\label{eq:wavemaker}
\end{equation}
and $\boldsymbol{x}_0$ denotes a position in space~\citep[see derivation of][]{Giannetti07}.

In cases where amplification is due to normal mechanisms, the resolvent modes can be used to find the wavemaker as long as they are normalized appropriately. The wavemaker approximates regions of the flow which are absolutely unstable or self-sustaining since perturbations are prevented from convecting due to reverse flow \cite{Juniper12}. \citet{Huerre85} have shown that when a mean profile of hyperbolic tangent form exhibits greater than 13.6\% reverse flow with respect to the free stream, the flow is absolutely unstable. The streamwise extent of absolute instability and the wavemaker is finite since flow reversal is confined to a certain portion of the flow. This information is encoded within the advection term $\overline{\boldsymbol{u}} \cdot \nabla()$ through the sign of $\overline{\boldsymbol{u}}$. We note that the overlap of the resolvent forcing and response modes was identified as a qualitative proxy for sensitivity for base flows by \citet{Brandt11}. Henceforth we will denote the wavemaker of Equation~\ref{eq:wavemaker} as the ``true'' wavemaker and the approximation using resolvent modes as the ``resolvent'' wavemaker. If the flow is convectively unstable, there is no region of reverse flow and so $\overline{\boldsymbol{u}}$ is always positive. In this case, the optimal response or stability mode will be downstream of the optimal forcing or adjoint mode. The inner product $| \hat{\boldsymbol{\phi}}_1^{*}\hat{\boldsymbol{\psi}}_1 |$ decreases when the modes are separated in space. \cite{Chomaz05} noted that this is due to the convective-type non-normality introduced from advection of the base flow.

\begin{figure}[!htb]
        \centering
        \subfloat[]{\includegraphics[scale=0.33]{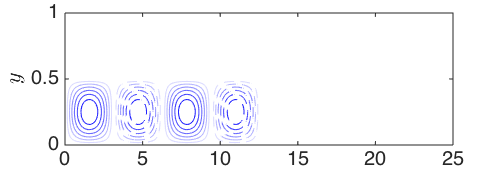}}
        \subfloat[]{\includegraphics[scale=0.33]{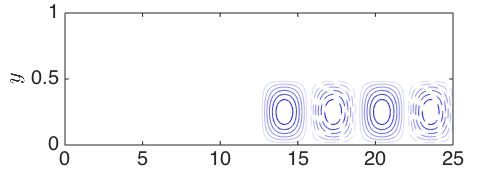}}

        \subfloat[]{\includegraphics[scale=0.33]{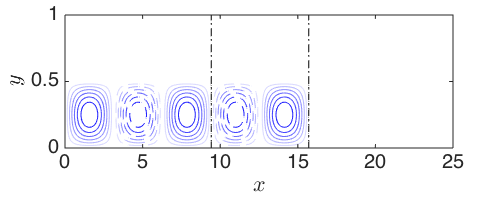}}
        \subfloat[]{\includegraphics[scale=0.33]{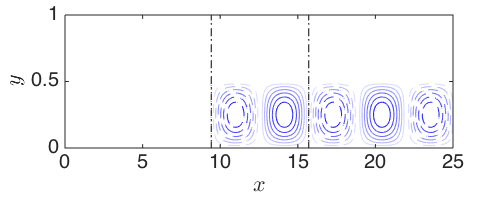}}

\caption{Cartoon of the leading forcing and response modes corresponding to the operator given in Equation~\ref{eq:L primitive} when the base or mean flow has variation in both directions. Panels (a, b) are the forcing and response, respectively, convective-type instability where mean flow advection results in the forcing (a) being upstream of the response (b). Panels (c, d) show the forcing and response, respectively, for an absolute-type instability. The vertical dashed-dotted lines denote the region of overlap in the streamwise direction of forcing (c) and response (d), i.e. the resolvent wavemaker.  Positive/negative isocontours are denoted by solid/dotted lines, respectively.}
\label{fig:cartoon_instab}
\end{figure}

The cartoon in Figure~\ref{fig:cartoon_instab} summarizes the relative positioning of the various modes for a 2D normal operator. The top row is characteristic of a convective-type instability where mean flow advection orients the optimal forcing upstream of a downstream, detached response. The bottom row represents a case where there is overlap between the forcing and response modes indicating a resolvent wavemaker where the flow exhibits an absolute-type instability. The implications of this approximate wavemaker region are discussed in the context of cylinder flow in Section~\ref{sec:cylinder}.


\subsection{Low-rank approximation of the resolvent through a dyad expansion}

We present what is believed to be a novel (if logical) expansion of the above analysis to demonstrate and formalize the conditions under which {analysis of the resolvent is likely to} identify stability modes as the most amplified disturbance. We also highlight when the rank-1 approximation is appropriate for amplifications which are normal in character.

As is more customary for the eigenvalue problem~\citep[][]{Luchini14, Schmid14} a dyad expansion of the resolvent $\boldsymbol{R}$ for a generic, non-singular linear operator $\boldsymbol{Q}$ can be performed,
\begin{equation}
\boldsymbol{R} = (z\boldsymbol{I}-\boldsymbol{Q})^{-1} = \sum_{j=1}^n \frac{1}{z-\lambda_j} \tilde{\boldsymbol{g}}_j\tilde{\boldsymbol{h}}_j^{*},
\label{eq:dyad Q}
\end{equation}
where $\tilde{\boldsymbol{g}}_j$ and $\tilde{\boldsymbol{h}}_j$ are the $j$th left and right eigenvectors of $\boldsymbol{Q}$, respectively. Since the objective of resolvent analysis is often the identification of the most amplified neutral disturbance, $i\omega$ is substituted for $z$ and the eigenvectors of the LNS operator for $\tilde{\boldsymbol{g}}_j$ and $\tilde{\boldsymbol{h}}_j$ into Equation~\ref{eq:dyad Q} to give
\begin{equation}
\mathcal{H}(\omega) = \sum^n_{j=1} \frac{1}{i\omega-\lambda_j}\tilde{\boldsymbol{u}}_j\tilde{\boldsymbol{v}}_j^{*}.
\label{eq:dyad H}
\end{equation}
Thus if the real part of an eigenvalue $\lambda_p$ is sufficiently close to zero and the forcing frequency $\omega$ is identical to the imaginary part, then its contribution to the series dominates over the contributions from all other eigenvalues. The resolvent, furthermore, can be approximated by the forward and adjoint eigenvectors corresponding to that frequency weighted by the inverse distance between the eigenvalue and the imaginary axis:

\begin{equation}
\mathcal{H}(\omega) \approx \frac{1}{i \omega - \lambda_p}\tilde{\boldsymbol{u}}_p\tilde{\boldsymbol{v}}_p^{*}.
\label{eq:dyad rank 1}
\end{equation}
Equation~\ref{eq:dyad rank 1} represents a rank-1 approximation of the resolvent operator using eigenvectors. In the context of base flows, the resolvent is singular at the critical Reynolds number since the real part of the least stable eigenvalue is identically zero when it crosses the imaginary axis. In the case of mean flows, which tend to be marginally stable \citep[e.g.][]{Reynolds67, Barkley06, Turton15}, Equation~\ref{eq:dyad rank 1} is applicable for eigenvalues near the imaginary axis. It is important to note that an eigenvalue does not have to be marginally stable, but it must be the dominant contribution to the series in Equation~\ref{eq:dyad H}. It is possible to obtain a rank-1 approximation of the resolvent even when an eigenvalue is highly damped as will be seen in Section~\ref{sec:cylinder}.

This rank-1 approximation fails when there is not sufficient separation of eigenvalues at the frequency of interest (see Appendix~\ref{sec:rank1} for an example). If there are several eigenvalues in the vicinity of the imaginary axis at a frequency $\omega$, then the resolvent operator can no longer be approximated by just one outer product in Equation~\ref{eq:dyad rank 1}. Equating the two low-rank approximations of the resolvent operator in terms of eigenvectors (Equation~\ref{eq:dyad rank 1}) and resolvent modes (Equation~\ref{eq:rank 1}) implies the following:

\begin{equation}
\sigma_1 \hat{\boldsymbol{\psi}}_1\hat{\boldsymbol{\phi}}_1^{*} \approx \frac{1}{i\omega - \lambda_r} \tilde{\boldsymbol{u}}_r \tilde{\boldsymbol{v}}^{*}_r \implies  \hat{\boldsymbol{\psi}}_1 \propto \ \tilde{\boldsymbol{u}}_r, ~~~ \hat{\boldsymbol{\phi}}_1 \propto \tilde{\boldsymbol{v}}_r,
\label{eq:proportional}
\end{equation}
since
\begin{equation}
\hat{\boldsymbol{\psi}}_1 \approx \frac{1}{\sigma_1 (i\omega - \lambda_r)}\tilde{\boldsymbol{u}}_r\tilde{\boldsymbol{v}}_r^{*} \hat{\boldsymbol{\phi}}_1 = \beta \tilde{\boldsymbol{u}}_r,
\label{eq:singular and eigenmodes}
\end{equation}
where $\beta$ is a complex constant. The leading resolvent response and forcing modes are proportional to the forward and adjoint eigenmodes, respectively, and this holds for any base or mean flow as long as only one eigenvalue leads to amplification. The similarity between the resolvent forcing and adjoint stability modes draws out how the resolvent operator contains sensitivity information, as described by,~\cite[e.g.][]{Qadri17}. The development is less amenable to non-normal mechanisms where the proximity of an eigenvalue to the imaginary axis does not necessarily govern the behavior of the resolvent (see, e.g. Figure~\ref{fig:pseudospectra}). Some guidelines for when a rank-1 approximation is expected to work for a general resolvent operator are presented in Section~\ref{sec:turbulence}.

\subsection{The relationship between spectral radius and spectral norm for approximately low-rank operators}

For a nonsingular linear operator $\boldsymbol{Q}$, we now seek to find an explicit relationship between the spectrum of $\boldsymbol{Q}$ and the spectral norm of its resolvent $\boldsymbol{R}(z) = (z\boldsymbol{I}-\boldsymbol{Q})^{-1}$. We are particularly interested in cases where $\boldsymbol{R}(z)$ is approximately low-rank (i.e., a small number of leading singular values are much larger than the others).

The spectral radius $\rho$ of an operator $\boldsymbol{Q}$ can be defined through the eigendecomposition $ \boldsymbol{V\Lambda V}^{-1}$
\begin{equation}
\rho(\boldsymbol{Q}) = \max_{\lambda_j\in \boldsymbol{\Lambda}}(|\lambda_j|).
\end{equation}
The spectral radius of the corresponding resolvent operator is
\begin{equation}
\rho(\boldsymbol{R}(z)) = \max_{\lambda_j\in \boldsymbol{\Lambda}}(|z - \lambda_j|^{-1})  =  \left[ \min_{\lambda_j\in \boldsymbol{ \Lambda}}(|z - \lambda_j|)\right]^{-1}.
\end{equation}
Note that, with this definition, Equation \ref{eq:resolvent eigenvalue} may be expressed as
$$  \rho(\boldsymbol{R}) \leq  \sigma_1 \leq \kappa   \rho(\boldsymbol{R}).  $$
For non-normal operators with large condition numbers, the upper and lower bounds span a large range, and thus do not give much insight into the size of the resolvent norm, $\sigma_1$.
To estimate the resolvent norm in terms of the spectral radius, we will make use of the relationship \cite{Gelfand41}
\begin{equation}
\label{eq:SpectralRadius}
\rho (\boldsymbol{R}) = \lim_{n\to\infty}\| \boldsymbol{R}^n \|^{1/n}.
\end{equation}
Suppose now that the largest singular value of resolvent operator (for a given $z$) is much larger than the rest, such that
$$\boldsymbol{R}(z) = \boldsymbol{\Psi}\boldsymbol{\Sigma}\boldsymbol{\Phi}^{*}
			     \approx\boldsymbol{\Psi}\boldsymbol{\Sigma}_1\boldsymbol{\Phi}^{*},
$$
where $\boldsymbol{\Sigma}_1$ is $\boldsymbol{\Sigma}$ with all but the first singular value set to zero.

Suppose in addition that this truncation is also accurate for powers of $\boldsymbol{R}$, i.e., that we have
\begin{equation}
\label{eq:PowerTrunc}
\boldsymbol{R}^n    \approx(\boldsymbol{\Psi}\boldsymbol{\Sigma}_1\boldsymbol{\Phi}^*)^n.
\end{equation}
% Note the differences in these assumptions.
Defining the quantity
$$r_{ij} = \frac{\boldsymbol{\phi}_i^*\boldsymbol{\psi}_j}{\boldsymbol{\phi}_j^*\boldsymbol{\psi}_j},$$
we then have
\begin{align*}
\left(\boldsymbol{\Phi}^{*} \boldsymbol{\Psi}\boldsymbol{\Sigma}_1\right)^n &
%= \sigma_1^n(\phi_1^*\psi_1)^{n-1}
%\begin{pmatrix}  \phi_1^*\psi_1 & 0 & \cdots & 0 \\
%\phi_2^*\psi_1 & 0 & \cdots & 0 \\
%\vdots & \vdots &\ddots & \vdots
%\end{pmatrix} \\
= \sigma_1^n(\boldsymbol{\phi}_1^*\boldsymbol{\psi}_1)^n
 \begin{pmatrix}  r_{11} & 0 & \cdots & 0 \\
r_{21} & 0 & \cdots & 0 \\
\vdots & \vdots &\ddots & \vdots
\end{pmatrix}.
\end{align*}
We may now estimate the norm of powers of the resolvent  as
%\begin{align*}
$$\| \boldsymbol{R}^n \| = \| \boldsymbol{\Phi}^{*}\boldsymbol{R}^n \boldsymbol{\Phi}\|  \\
			\approx \|  \left(\boldsymbol{\Phi}^{*} \boldsymbol{\Psi}\boldsymbol{\Sigma}_1\right)^n\| \\
			 = \sigma_1^n|\boldsymbol{\phi}_1^*\boldsymbol{\psi}_1|^{n-1}\| \boldsymbol{r}\|,
			 $$
where $  \boldsymbol{r} = [r_{11} \ r_{21} \ \cdots]^T$, and we have used the fact that $\boldsymbol{\Phi}$ is unitary.
Consequently, assuming that Equation \ref{eq:PowerTrunc} holds, Equation \ref{eq:SpectralRadius} results in the estimate
\begin{equation}
\rho(\boldsymbol{R}) \approx \sigma_1 |\boldsymbol{\phi}_1^*\boldsymbol{\psi}_1 |.
\end{equation}
In other words, we may estimate that the resolvent norm is larger than the lower bound in Equation \ref{eq:resolvent eigenvalue} by a factor of $|\boldsymbol{\phi}_1^*\boldsymbol{\psi}_1 |^{-1}$. This analysis relied on the rather restrictive assumption that only the leading singular value was large. If there is a pair of large singular values, as is often the case in channel flows (owing to spatial symmetry across the mid-plane of the channel) then we may generalize the argument as follows.  Suppose that $\sigma_1$ and $\sigma_2$ are of comparable size, and that all other singular values are negligibly small. If we further assume that $|\boldsymbol{\phi}_1^*\boldsymbol{\psi}_2|, \ |\boldsymbol{\phi}_2^*\boldsymbol{\psi}_1| \approx 0$, then we find that
\begin{align*}
\left(\boldsymbol{\Phi}^{*} \boldsymbol{\Psi}\boldsymbol{\Sigma}\right)^n &\approx
 \begin{pmatrix} \sigma_1\boldsymbol{\phi}_1^*\boldsymbol{\psi}_1  & \sigma_2 \boldsymbol{\phi}_1^*\boldsymbol{\psi}_2 & 0 & \cdots & 0 \\
 \sigma_1\boldsymbol{\phi}_2^*\boldsymbol{\psi}_1 & \sigma_2\boldsymbol{\phi}_2^*\boldsymbol{\psi}_2 &  0 & \cdots & 0 \\
\vdots & \vdots  &  \vdots &\ddots & \vdots
\end{pmatrix}^n \\
&\approx \sigma_1^n  |\boldsymbol{\phi}_1^*\boldsymbol{\psi}_1|^{n-1}
 \begin{pmatrix}  r_{11} & 0 & \cdots & 0 \\
r_{21} & 0 & \cdots & 0 \\
\vdots & \vdots &\ddots & \vdots
\end{pmatrix} +
 \sigma_2^n  |\boldsymbol{\phi}_2^*\boldsymbol{\psi}_2|^{n-1}
 \begin{pmatrix}  0 & r_{12} & 0 & \cdots & 0 \\
0 & r_{22} & 0 & \cdots & 0 \\
\vdots & \vdots & \vdots &\ddots & \vdots
\end{pmatrix},
\end{align*}
which, following the same approach as before, gives
\begin{equation}
\rho(\boldsymbol{R}) \approx \max\{\sigma_1 |\boldsymbol{\phi}_1^*\boldsymbol{\psi}_1 |,\sigma_2 |\boldsymbol{\phi}_2^*\boldsymbol{\psi}_2 |\}.
\end{equation}

Thus the inverse of $| \hat{\boldsymbol{\phi}}_1^{*}\hat{\boldsymbol{\psi}}_1 |$ can be interpreted as the contribution of non-normality to the resolvent norm. We can also identify the product $\sigma_1 |i\omega - \lambda|$ as a quantification of non-normality since $|i \omega - \lambda |^{-1}$ represents the resonance contribution to the resolvent norm. However, since highly amplified modes may occur at non-resonant frequencies, the contribution from $| i \omega - \lambda |$ is typically overestimated as it is likely for a pseudoeigenvalue to reside much closer to the imaginary axis than the nearest eigenvalue of the unperturbed spectrum. As will be seen in Sections~\ref{sec:cylinder} and~\ref{sec:turbulence}, these two predictions tend to agree in cases where amplification can be attributed to a single eigenvalue and mean stability analysis is valid.

Having examined the implications of the structure of the operator on amplification and forcing and response modes, we now consider two real example flows. Low Reynolds number cylinder flow is used to investigate the choice of base or mean flow as the linear stability threshold is crossed (Section~\ref{sec:cylinder}). A canonical wall turbulence configuration is employed to identify the influence of the various terms in the resolvent on the resulting SVD (Section~\ref{sec:turbulence}).

%%%%%%%%%%%%%%%%%%%%%%%%%%%%%%%%%%%%%%%%%%%%%%%%%%%%

\section{Application to cylinder flow}
\label{sec:cylinder}

We apply a global resolvent analysis to the base and mean velocity profiles for cylinder flows under the critical Reynolds number $Re_c \leq 47$ \citep{Provansal87, Sreenivasan87, Noack94}, as well as mean flows of the 2D laminar vortex shedding regime where $Re\leq 189$ \citep{Barkley96}. Cylinder flow is a particularly suitable choice to investigate trends associated with the wavemaker since it exhibits a region of absolute instability.

\subsection{Numerical methods}
The relevant procedures for computing the two-dimensional base and mean flows, $\boldsymbol{U}_0$ and $\overline{\boldsymbol{u}}$, are detailed here before applying the analysis tools.

The NSE (Equation~\ref{eq:NS}) are non-dimensionalized by the cylinder diameter $D$ and inlet velocity $U_{\infty}$ which are both set to unity. For the base flow calculation, a uniform inlet velocity condition is prescribed while no-slip Dirichlet boundary conditions are applied to the cylinder surface, symmetric conditions to the upper and lower boundaries, and advective conditions to the outlet. The nonlinear equations for $\boldsymbol{U}_0$ are solved using a Newton method on a finite-element mesh generated by FreeFem++ (see \cite{Hecht12}). Taylor-Hood finite elements (P1b, P1b, P1 for $U_0$, $V_0$, and $P_0$ respectively) are used for spatial discretizations. The computational domain $\Omega$ spans $-30 \leq x/D \leq 60, -25 \leq y/D \leq 25$ with the cylinder centered at the origin and the mesh is made up of 104,214 triangles resulting in 365,358 degrees of freedom for velocity and pressure.

A DNS of the cylinder flow is also performed to obtain the mean flow profile using FreeFem++ with the same boundary conditions and mesh. A second-order semi-implicit time discretization is employed with a non-dimensional time step $\Delta t = 0.02$. Beyond $Re_c$, the simulated flow settles into regular vortex shedding at a fixed amplitude $A$ and temporal frequency $\omega_s$ where the subscript $s$ denotes shedding. The mean flow $\overline{\boldsymbol{u}}$ is computed by time-averaging the DNS state vector over 25 complete shedding cycles. The linear operators are formed in FreeFem++ and the only boundary condition which differs with respect to the base flow calculation is at the inlet where homogeneous boundary conditions are enforced so that the perturbations vanish at infinity. The eigenvalues are computed using a shift-and-invert strategy, the details of which are discussed in~\citet{Nayar93}. The generalized eigenvalue problem is then solved with the Implicitly Restarted Arnoldi method using the ARPACK library developed by~\citet{Lehoucq96}.

The singular values of the resolvent operator are computed in a manner outlined by~\citet{Sipp13}; a brief summary of the procedure is presented here. The singular value problem is reformulated as the following eigenvalue problem:
\begin{equation}
\mathcal{H}(\omega)^{*}\mathcal{H}(\omega)\hat{\boldsymbol{\phi}_i} = \sigma_i^2 \hat{\boldsymbol{\phi}_i},
\label{eq:resolvent computation}
\end{equation}
where $\hat{\boldsymbol{\phi}_i}$ is the $i$th right singular vector corresponding to the singular value $\sigma_i$ of $\mathcal{H}(\omega)$. The largest eigenvalues of the Hermitian operator $\mathcal{H}(\omega)^*\mathcal{H}(\omega)$ are computed using the ARPACK library and the parallel MUMPS solver developed by~\cite{Amestoy01}. The response modes are then computed from Equation~\ref{eq:resolvent operator}.

\subsection{Base flow velocity profile}

\begin{figure}
\centering
        \subfloat[]{\includegraphics[scale=0.4]{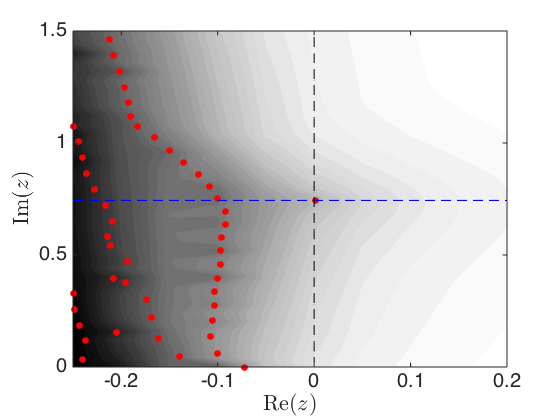}}
        \subfloat[]{\includegraphics[scale=0.4]{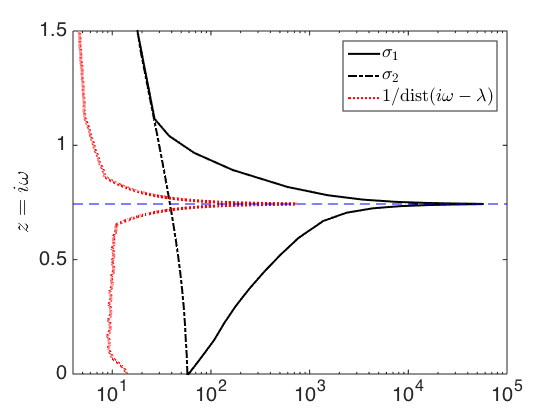}}
\caption{(a) Spectrum (red dots) and pseudospectrum (filled contours, $\epsilon$ increasing as colors change from dark to light) of the LNS operator for the cylinder base flow at $Re = 47$. (b) The resolvent norm, $\sigma_1$ (solid line), i.e. the value of $\epsilon^{-1}$ along the imaginary axis, second largest singular value $\sigma_2$ (dash-dotted line) and inverse distance from the imaginary axis to the nearest eigenvalue (dotted red line).}
\label{fig:svalsbase}
\end{figure}

A resolvent analysis is performed on the cylinder base flow for various Reynolds numbers over a range of $\omega$. Contours of the pseudospectrum for $Re = 47$ are overlaid onto the spectrum, which is in agreement with~\citet{Sipp07} to within the sensitivity to the mesh geometry, in Figure~\ref{fig:svalsbase}(a). The variation of the resolvent norm along the imaginary axis, i.e. $\sigma_1$, is plotted alongside the second singular value of the resolvent, $\sigma_2$, and the inverse of the distance between the nearest eigenvalue and the imaginary axis in Figure~\ref{fig:svalsbase}(b). There is only one frequency $\omega_{\text{max}}$ of interest, where the first singular value is several orders of magnitude larger than all the others (only two are shown for clarity). For the cylinder base flow at $Re = 47$, $\omega_{\text{max}} = 0.742$ corresponds to the true $\omega_s$ at the onset of vortex shedding. The amplification is significantly lower for all other frequencies, including harmonics, and $\omega_{\text{max}}$ is very close to the imaginary part of the least stable eigenvalue. In short, resolvent analysis of the base flow yields useful predictions for the frequency of the unsteady flow in the vicinity of the bifurcation like its stability counterpart.

\begin{figure}
        \centering
        \subfloat[]{\includegraphics[scale=0.4,trim=0.5cm 3cm 0.5cm 3cm, clip=true]{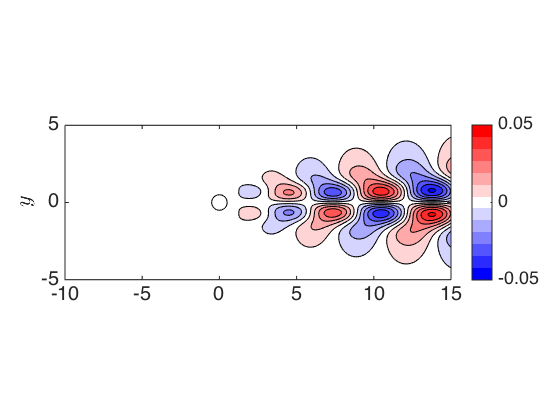}}
        \subfloat[]{\includegraphics[scale=0.4,trim=0.5cm 3cm 1cm 3cm, clip=true]{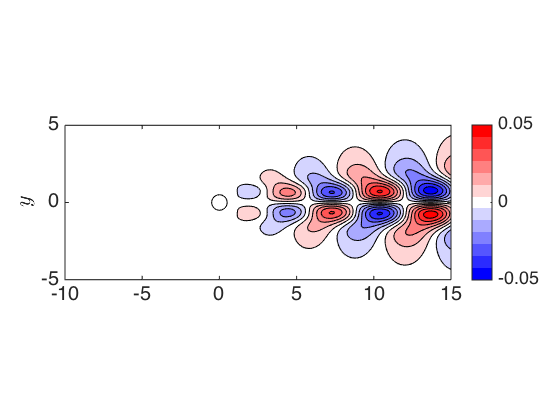}}

        \subfloat[]{\includegraphics[scale=0.4,trim=0.5cm 3cm 1cm 3cm, clip=true]{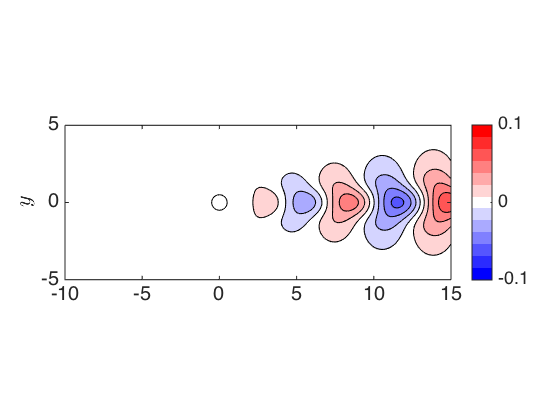}}
	\subfloat[]{\includegraphics[scale=0.4,trim=0.5cm 3cm 1cm 3cm, clip=true]{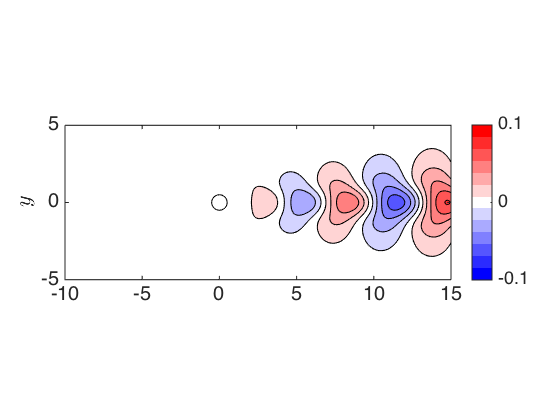}}

	\subfloat[]{\includegraphics[scale=0.4,trim=0.5cm 3cm 1cm 3cm, clip=true]{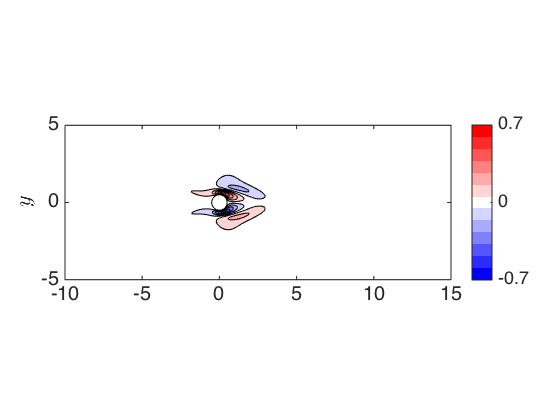}}
	\subfloat[]{\includegraphics[scale=0.4,trim=0.5cm 3cm 1cm 3cm, clip=true]{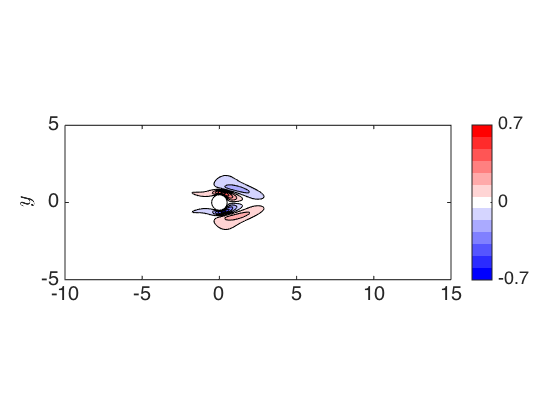}}

	\subfloat[]{\includegraphics[scale=0.4,trim=0.5cm 3cm 1cm 3cm, clip=true]{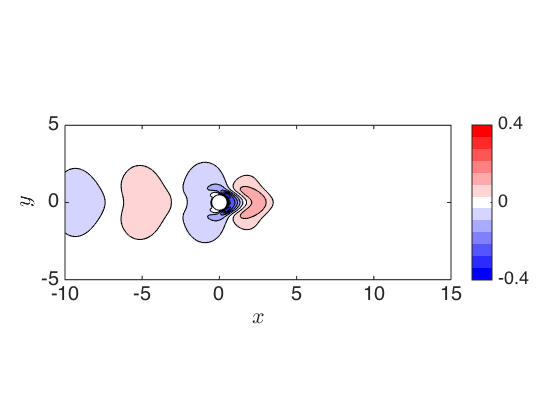}}
	\subfloat[]{\includegraphics[scale=0.4,trim=0.5cm 3cm 1cm 3cm, clip=true]{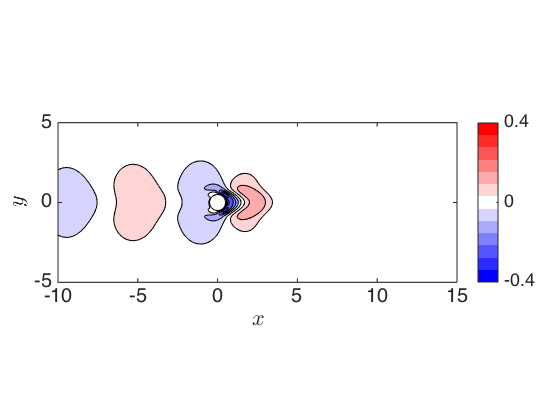}}

\caption{Comparison of stability modes (left) with resolvent modes (right) at the critical Reynolds number $Re_c = 47$ and a temporal frequency of $\omega = 0.742$. Panels (a, b) are the streamwise component of the forward or response mode, (c, d) are the transverse component of the forward or response mode, (e, f) are the streamwise component of the adjoint or forcing mode, and (g, h) are the transverse component of the adjoint or forcing mode. The eigenmodes and resolvent mode shapes are essentially indistinguishable for this flow.}\label{fig:comparison_baseflows}
\end{figure}

The previous observations reinforce why stability analysis about the base flow can predict $Re_c$. Only one structure at the globally most amplified frequency is prone to significant amplification at subcritical Reynolds numbers and it is the first to become unstable. This is characteristic of an absolute instability mechanism in which frequency selection is not influenced by background noise. The stability modes and resolvent modes are nearly identical as seen in Figure~\ref{fig:comparison_baseflows}.
The effect of lift-up is weak since the energy is fairly evenly distributed in the $u-$ and $v-$components of both the forcing and response modes in Figure~\ref{fig:comparison_baseflows}. Mean flow advection, on the other hand, plays a significant role in the spatial support of the forcing and response modes which are located upstream and downstream of the cylinder.~\citet{Chomaz05} made an analogous observation for the forward and adjoint eigenmodes and attributed this to convective non-normality.  Since the resolvent operator is low-rank, computing $ | \hat{\boldsymbol{\phi}}_1^{*}\hat{\boldsymbol{\psi}}_1 |^{-1}$ is a good estimate of the non-normal amplification experienced by the flow. We obtain a value of $\| \hat{\boldsymbol{\phi}}_1^{*} \hat{\boldsymbol{\psi}}_1 \|^{-1} = 79.4$ which is in good agreement with $\sigma_1(\omega_{\text{max}})| i\omega_{\text{max}}-\lambda_{\text{ls}}| = 79.3$ (see Table~\ref{tab:mode comparison}) in Figure~\ref{fig:svalsbase}(b). There is also a physical overlap of the forcing and response modes in the region where the flow is absolutely unstable, which can be attributed to the extent of reverse flow.

\begin{figure}
        \centering
        \subfloat[]{\includegraphics[scale=0.27,trim=0cm 3cm 0cm 3.5cm, clip=true]{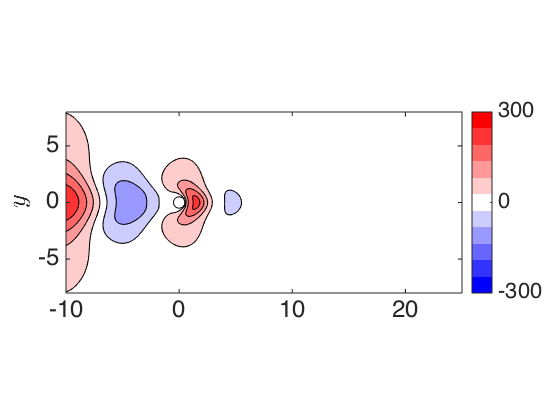}}
        \subfloat[]{\includegraphics[scale=0.27,trim=0cm 3cm 0cm 3.5cm, clip=true]{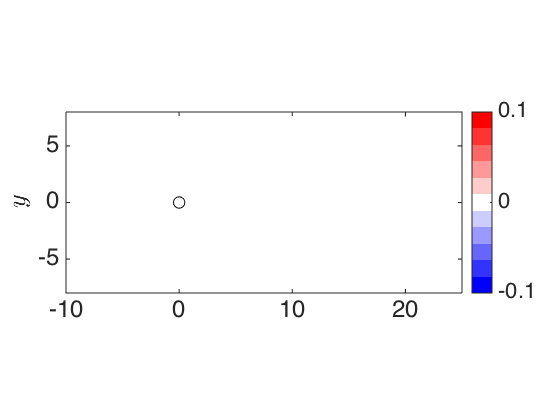}}
        \subfloat[]{\includegraphics[scale=0.27,trim=0cm 3cm 0cm 3.5cm, clip=true]{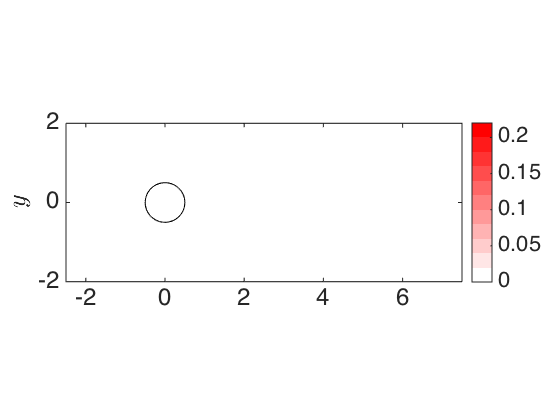}}
	
	\subfloat[]{\includegraphics[scale=0.27,trim=0cm 3cm 0cm 3.5cm, clip=true]{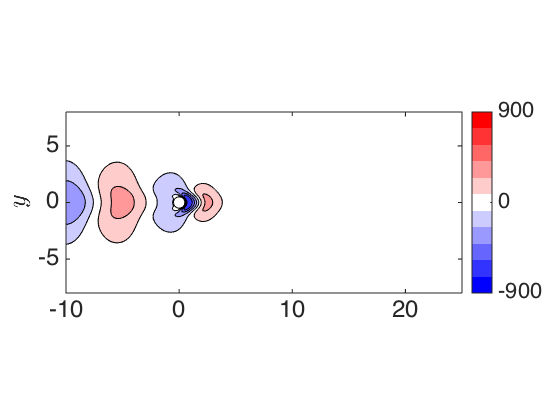}}
        \subfloat[]{\includegraphics[scale=0.27,trim=0cm 3cm 0cm 3.5cm, clip=true]{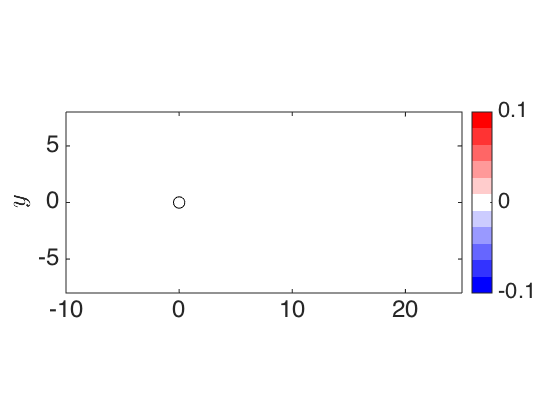}}
        \subfloat[]{\includegraphics[scale=0.27,trim=0cm 3cm 0cm 3.5cm, clip=true]{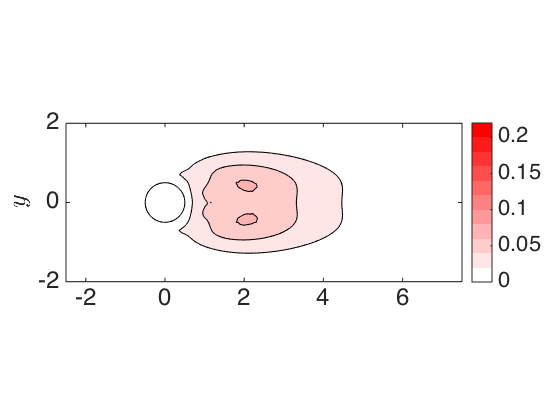}}

	\subfloat[]{\includegraphics[scale=0.27,trim=0cm 3cm 0cm 3.5cm, clip=true]{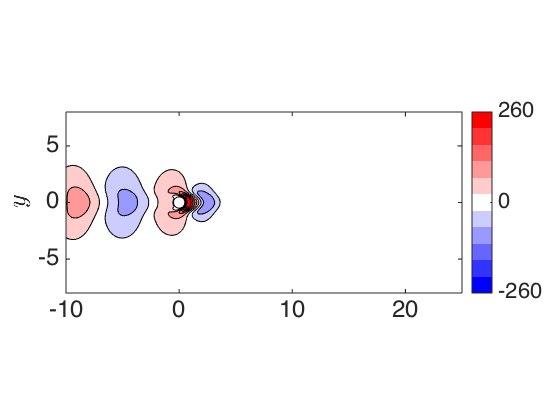}}
        \subfloat[]{\includegraphics[scale=0.27,trim=0cm 3cm 0cm 3.5cm, clip=true]{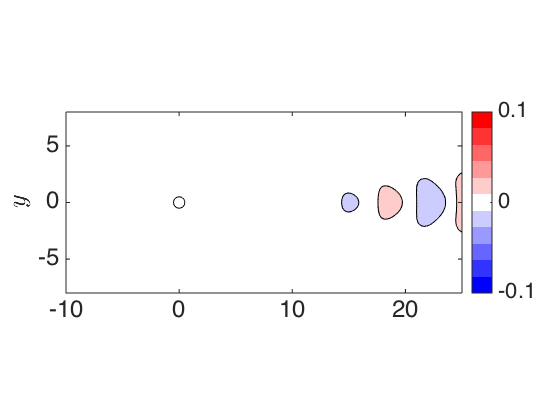}}
        \subfloat[]{\includegraphics[scale=0.27,trim=0cm 3cm 0cm 3.5cm, clip=true]{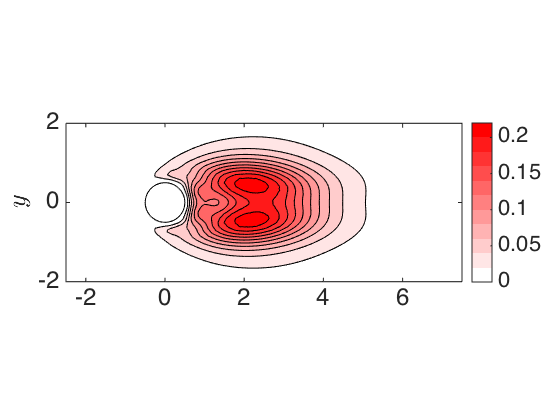}}
	
	\subfloat[]{\includegraphics[scale=0.27,trim=0cm 3cm 0cm 3.5cm, clip=true]{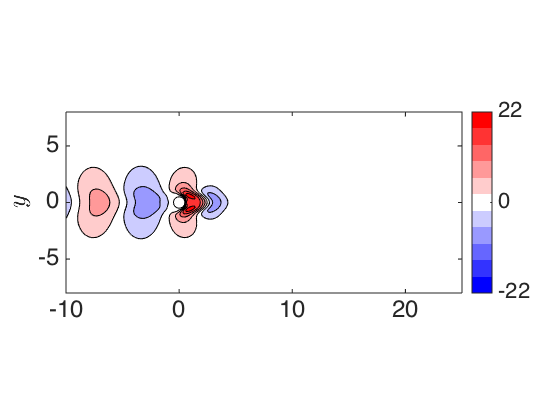}}
        \subfloat[]{\includegraphics[scale=0.27,trim=0cm 3cm 0cm 3.5cm, clip=true]{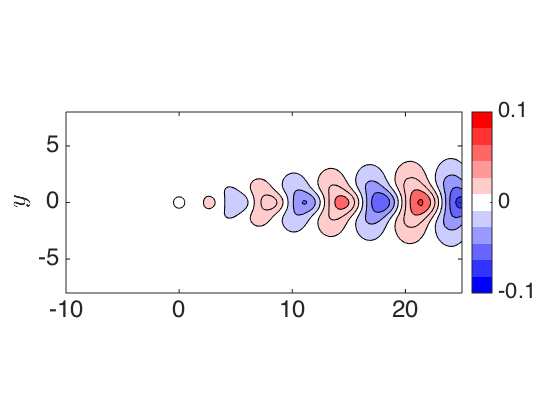}}
        \subfloat[]{\includegraphics[scale=0.27,trim=0cm 3cm 0cm 3.5cm, clip=true]{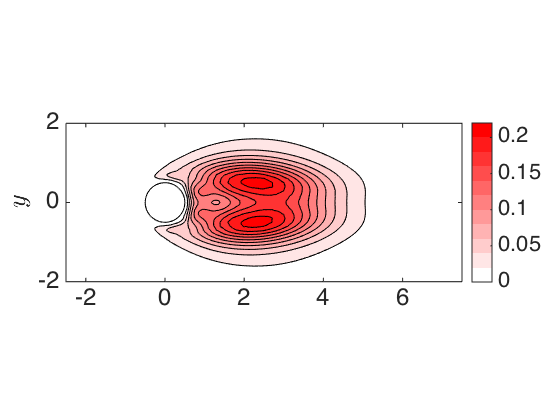}}
	
	\subfloat[]{\includegraphics[scale=0.27,trim=0cm 2.5cm 0cm 3.5cm, clip=true]{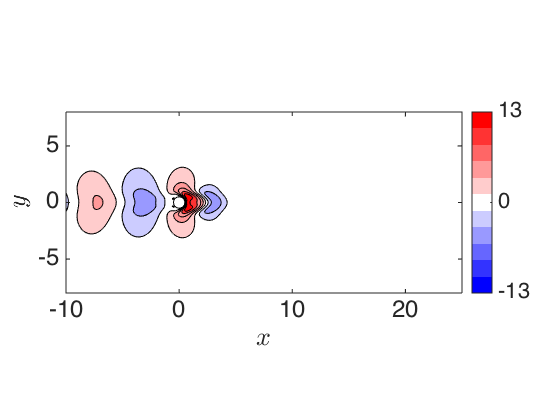}}
        \subfloat[]{\includegraphics[scale=0.27,trim=0cm 2.5cm 0cm 3.5cm, clip=true]{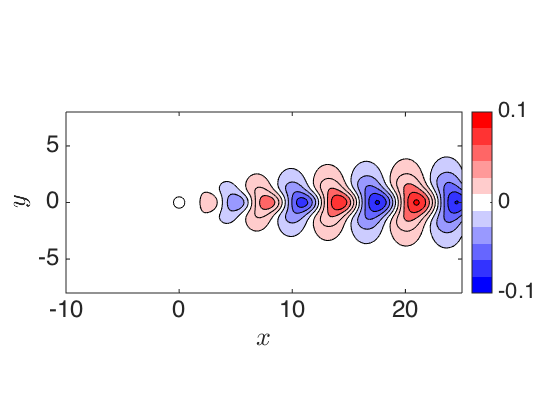}}
        \subfloat[]{\includegraphics[scale=0.27,trim=0cm 2.5cm 0cm 3.5cm, clip=true]{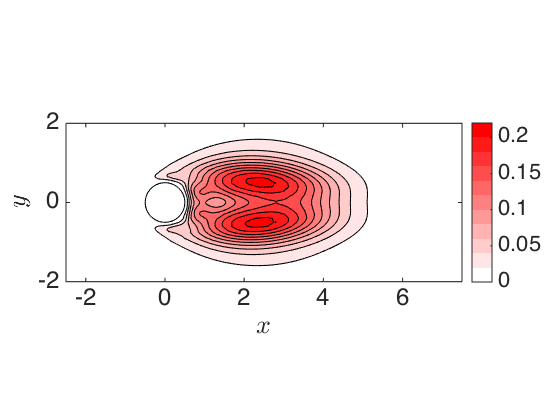}}

\caption{Contours of the transverse velocity for the leading adjoint modes $\tilde{v}^{\dagger}$ (left) and forward modes $\tilde{v}$ (middle) of the base flow. The Reynolds numbers ($Re =$15; 25; 35; 45; and 50) increase from top to bottom. The wavemaker $\mathcal{W}$ (right) is computed using the forward and adjoint modes. Contour levels are not identical for the adjoint modes which are normalized based on the forward modes. Note that the streamwise velocity component has not been plotted even though the wavemaker depends on this quantity.}\label{fig:wavemaker_baseflows}
\end{figure}

The least stable global mode and its adjoint counterpart are computed for various Reynolds numbers near and below $Re_c$ to illustrate the cylinder transition from convective to absolute instability. Figure~\ref{fig:wavemaker_baseflows} juxtaposes the $v$-component of the adjoint mode, forward mode, and wavemaker. The forward mode has unit magnitude while the adjoint has been normalized with respect to the forward mode such that their inner product is unity. A wavemaker first appears for $Re = 25$, the Reynolds number at which~\citet{Monkewitz88} determined the cylinder wake is absolutely unstable. There is no wavemaker for lower Reynolds numbers due to the downstream location of the forward eigenmode which is a consequence of mean flow advection; the strength of the reverse flow is not sufficient to produce an overlap region. For the lowest two Reynolds numbers considered in Figure~\ref{fig:wavemaker_baseflows}, the contour levels of the forward eigenmode immediately behind the cylinder are three orders of magnitude smaller than their higher Reynolds number counterparts. The downstream location where the contour levels are significant does not appear within the plotted domain. As the Reynolds number increases, the velocity deficit grows  and the reverse flow directly behind the cylinder strengthens. The forward eigenmode gradually appears closer to the cylinder until there is a nontrivial overlap between it and its adjoint counterpart.

Beyond the critical Reynolds number, the region of the flow which is absolutely unstable is sufficiently long for the flow to become globally unstable. Perturbations grow exponentially in time until they are saturated by nonlinearities. The resulting velocity fluctuations are dominated by the vortex shedding. Once the flow has reached a limit cycle, the shedding frequency is different from that predicted by resolvent analysis of the base flow since the frequency of the least stable perturbations is altered during the saturation process. Additionally, the mean recirculation region behind the cylinder is shorter than its base flow counterpart in the streamwise direction.

\subsection{Mean velocity profile}

The focus of resolvent analysis typically shifts when using the mean velocity profile rather than the base flow. The goal becomes identification of the energetically important structures and their frequencies in the unsteady flow rather than prediction of the external forcing and structure which appears when the flow becomes unsteady.

\begin{figure}
\centering
        \subfloat[]{\includegraphics[scale=0.4]{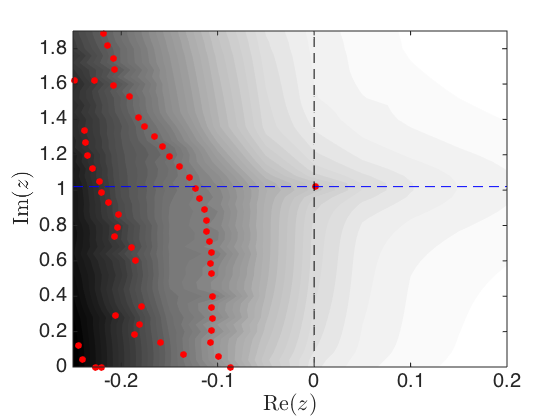}}
        \subfloat[]{\includegraphics[scale=0.4]{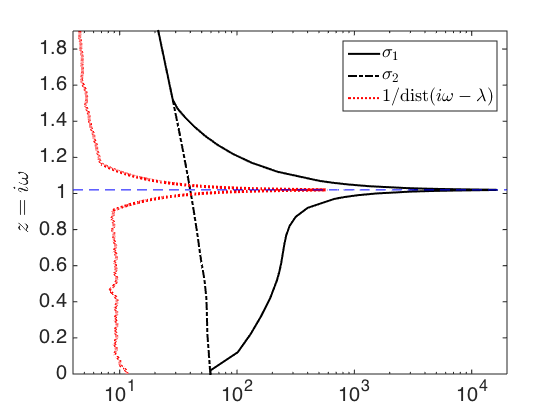}}
\caption{(a) Spectrum (red dots) and pseudospectrum (filled contours, $\epsilon$ increasing as colors change from dark to light) of the LNS operator for the cylinder mean flow at $Re = 100$. (b) The resolvent norm, $\sigma_1$ (solid line), i.e. the value of $\epsilon^{-1}$ along the imaginary axis, second largest singular value $\sigma_2$ (dash-dotted line) and inverse distance from the imaginary axis to the nearest eigenvalue (dotted red line).}
\label{fig:svalsmean}
\end{figure}

\begin{figure}
        \centering
        \subfloat[]{\includegraphics[scale=0.4]{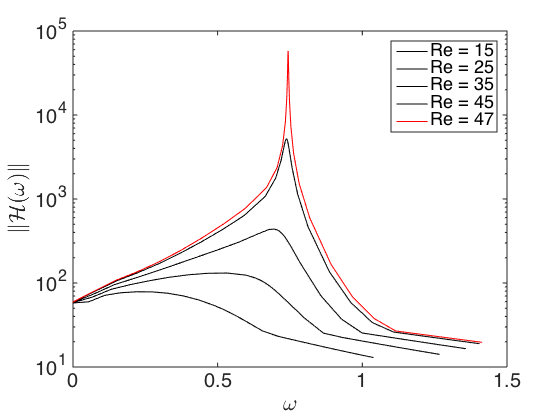}}
        \subfloat[]{\includegraphics[scale=0.4]{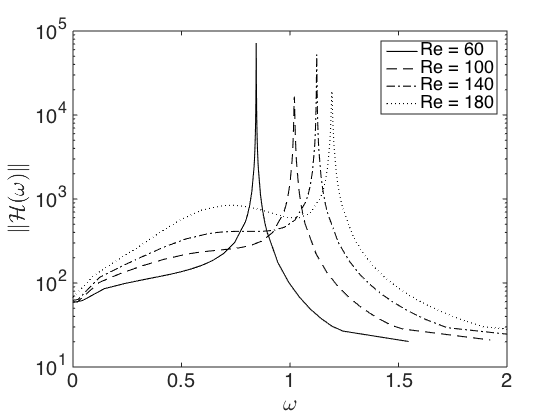}}

\caption{(a) The resolvent norm for the critical (solid red line), subcritical (solid black lines), and supercritical (dotted black lines) base flows. (b) The resolvent norm for supercritical mean flows.}\label{fig:svalscomp}
\end{figure}

Contours of the pseudospectrum corresponding to the mean flow in the 2D laminar shedding regime are overlaid with the spectrum of the mean flow at $Re = 100$ in Figure~\ref{fig:svalsmean}. The resolvent norm along the imaginary axis is also plotted alongside the second largest singular value. Similar to the base flow resolvent, there is only one frequency at which there is a resonant peak.
Unlike the base flow case, the most amplified frequency at supercritical Reynolds numbers correctly predicts the shedding frequency as seen in Figure~\ref{fig:svalscomp}. While the resolvent norm always peaks at a distinct frequency for all cases, the growth rate of the least stable eigenvalue of the base flow continues to grow while the frequency remains roughly constant. Figure~\ref{fig:svalscomp} shows that the largest peak occurs at the stability limit, $Re_c$. The maximum amplification, which here is proportional to the inverse distance between the eigenvalue and the imaginary axis, indicates the progression of the least stable pole across the complex plane and over the imaginary axis. The resolvent norm has not been plotted for supercritical base flows since the resolvent attempts to quantify the size of perturbation necessary for the spectrum to cross the neutral axis. For the base flow at the critical Reynolds number $Re_c$ and mean flows where $Re > Re_c$, the size of this perturbation is very small leading to very highly amplified disturbances.

The peak resolvent norm for the mean flows has no discernible pattern in Figure~\ref{fig:svalscomp} since the real part of the eigenvalue is approximately zero. It is very sensitive, therefore, to the spatial resolution and temporal convergence of the mean flow in addition to the discretization of $\omega$. Nevertheless, proportionality between the resolvent and stability mode shapes can be expected. Substituting $i\omega_s$ for $\lambda$ into Equation~\ref{eq:dyad H}, since the real part of the marginally stable mode is nearly zero, yields

\begin{equation}
\mathcal{H}(\omega_s) \approx \tilde{\boldsymbol{u}}_s \tilde{\boldsymbol{v}}_s^{*} \approx \hat{\boldsymbol{\psi}}_s\hat{\boldsymbol{\phi}}_s^{*}.
\label{eq: mean resolvent stability}
\end{equation}
The sum in Equation~\ref{eq:dyad H} is dominated by the contribution from the marginally stable mode so the resolvent operator can be approximated by the outer product of the marginally stable mode and its adjoint counterpart or the optimal resolvent response and forcing modes at $\omega_s$. Similar to the base flow case, we can quantify $| \hat{\boldsymbol{\phi}}_1^{*}\hat{\boldsymbol{\psi}}_1 |^{-1} = 26.9$ and this agrees fairly well with the ratio of $\sigma_1(\omega_{s})| i\omega_{s}-\lambda_{\text{s}}| = 28.9$ (see Table~\ref{tab:mode comparison}).

\begin{figure}
        \centering
        \subfloat[]{\includegraphics[scale=0.4,trim=0.5cm 3cm 1cm 3cm, clip=true]{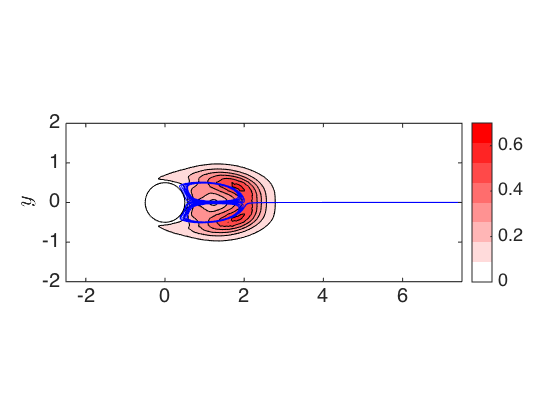}}
        \subfloat[]{\includegraphics[scale=0.4,trim=0.5cm 3cm 1cm 3cm, clip=true]{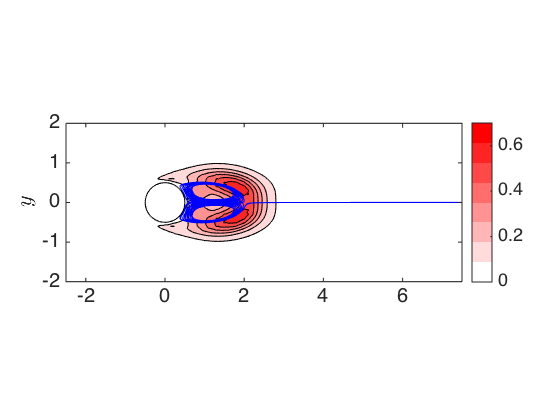}}

        \subfloat[]{\includegraphics[scale=0.4,trim=0.5cm 3cm 1cm 3cm, clip=true]{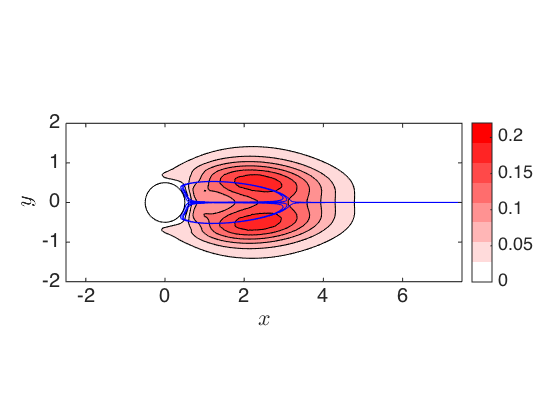}}
        \subfloat[]{\includegraphics[scale=0.4,trim=0.5cm 3cm 1cm 3cm, clip=true]{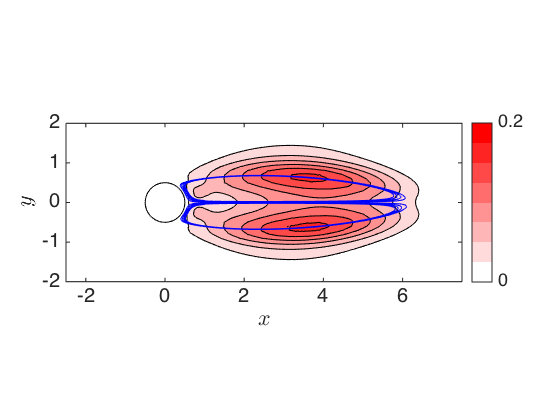}}
\caption{Wavemakers for mean flow at $Re = 100$ computed from stability modes (a) and resolvent modes (b). Wavemakers for the base flow at $Re = 47$ (c) and $Re = 100$ (d). Blue lines superimpose the mean flow streamlines which delineate the mean recirculation bubble.}\label{fig:wavemakers}
\end{figure}

Rather than comparing the stability and resolvent mode shapes as shown in Figure~\ref{fig:comparison_baseflows}, the true and approximated wavemakers associated with the mean flow, computed using stability and resolvent modes, respectively, are compared for $Re=100$ in Figure~\ref{fig:wavemakers}(a,b). The stability wavemaker is in good agreement with the one computed by~\citet{Meliga16}. A very close match between the extent of the two wavemaker regions can be observed, implying that the underlying modes are indeed proportional to each other. Streamlines from the mean flow are superimposed to observe how the wavemaker is related to the mean recirculation bubble, the size of which depends on Reynolds number. The length of the recirculation bubble scales with the streamwise extent of the wavemaker region for any Reynolds number for either the mean flow as seen in Figure~\ref{fig:wavemakers}(a,b) or base flow as seen in Figure~\ref{fig:wavemakers}(c,d). The advection term in the operator $\overline{\boldsymbol{u}} \cdot \nabla ()$ changes sign in the reverse flow region behind the cylinder which then guarantees overlap between the adjoint and forward eigenvectors, per Section~\ref{ssec:meanadvection}.

The true (stability) wavemaker regions associated with the base profile are shown in Figures~\ref{fig:wavemakers}(c) and (d) for $Re = 47$ and $Re=100$, respectively. Figures~\ref{fig:wavemakers}(a) and (d) compare the mean and base wavemakers at $Re = 100$. The main difference between base and mean flows is that as the Reynolds number increases, both the mean recirculation bubble~\cite{Zielinska97} and wavemaker region shrink, whereas for the unstable base flows, increasing the Reynolds number will also increase the streamwise length of the recirculation bubble and wavemaker region (Figures~\ref{fig:wavemakers}(c)-(d)). Therefore, the wavemaker computed from the mean flow is a better indication of the region of instability in the flow as it is responsible for the selection of the nonlinear frequency and amplitude of the limit cycle.

\subsection{Resolvent modeling of (low Reynolds number) cylinder flow}

In the case of flow around a circular cylinder, the resolvent identifies only stability mechanisms for both the base and mean flow cases. There are no frequencies at which amplification occurs due to pseudoresonance. This is consistent with the work of~\citet{Abdessemed09} who investigated direct transient growth analysis to study its role in the primary and secondary bifurcations of cylinder flow. Since only one mode becomes unstable, the effect of the nonlinearities is to saturate the growth mechanism and this alters the frequency of the structure \citep{Barkley06}. While the base flow is unable to predict the nonlinear frequency in the saturated state,  it can be correctly predicted from the mean flow. Thus in the case of stability mechanisms leading to unsteadiness (e.g. cylinder flow, Rayleigh-B\'enard convection~\citep{Turton15}), mean stability analysis is successful at predicting the frequency of the unsteady flow.

%%%%%%%%%%%%%%%%%%%%%%%%%%%%%%%%%%%%%%%%%%%%%%%%%%%%%%%%%

\section{Application to wall turbulence}
\label{sec:turbulence}

We now consider turbulent channel flow in the context of the discussion from Section~\ref{sec:amp}. Unlike the cylinder flow, which is an oscillator with intrinsic dynamics that are insensitive to background noise, this flow is an example of a noise-amplifier; as such, pseudoresonance plays a big role and leads to significant amplification at non-resonant frequencies. Due to its geometric simplicity, we choose channel flow at $Re_{\tau} = 2000$ which has a parallel mean velocity profile $\overline{\boldsymbol{u}} = \overline{u}(y)$.

\subsection{Numerical Methods}

We again consider the non-dimensional, incompressible NSE where the channel half-height $h$ and the friction velocity $u_{\tau} = \sqrt{\tau_w/\rho}$ (where $\tau_{w}$ is the wall shear stress, $\rho$ is the density) are the characteristic scales to obtain
\begin{subequations}\label{eq:NSE}
\begin{align}
\partial_t \boldsymbol{u} + \boldsymbol{u} \cdot \nabla \boldsymbol{u} &= -\nabla p + {Re_\tau}^{-1} \nabla^2 \boldsymbol{u} \\
\label{eq:Continuity}
\nabla \cdot \boldsymbol{u} &= 0,
\end{align}
\end{subequations}
where $Re_\tau = hu_{\tau}/\nu$ and $\nu$ is the kinematic viscosity. The streamwise direction and spanwise directions are periodic, and the wall-normal domain extends from $y/h =-1$ to $y/h=1$ with no-slip and no-penetration conditions imposed at the wall. The velocity field is Reynolds decomposed into the sum of a spatio-temporal mean and fluctuations,
\begin{equation}\label{eq:RD}
\boldsymbol{u}(x,y,z,t) = \overline{u}(y) + \boldsymbol{u'}(x,y,z,t).
\end{equation}
Here we assume the mean velocity profile is known a priori from an eddy viscosity model \cite{Reynolds67} as discussed in \cite{Moarref13}. We express the fluctuations as Fourier modes in the streamwise/spanwise directions and in time,
\begin{equation}\label{eq:FT}
\hat{\boldsymbol{u}}(k_x,k_z,\omega;y) = \displaystyle\int_{-\infty}^\infty \displaystyle\int_{-\infty}^\infty \displaystyle\int_{-\infty}^\infty \boldsymbol{u}'(x,y,z,t)e^{-i(k_xx + k_zz -\omega t)}\mathrm{d}x \mathrm{d}z \mathrm{d}t,
\end{equation}
where $k_x$ is the streamwise wavenumber, $ k_z $ is the spanwise wavenumber, and $\omega$ is the radial frequency. Upon elimination of the pressure term, we can express the governing equations in terms of the fluctuating vertical velocity $ \hat{v} $ and normal vorticity $ \hat{\eta} = ik_z \hat{u} - ik_x \hat{w}$,
\begin{equation}\label{eq:OS_SQ}
-i\omega
\left(
\begin{array}{c}
\hat{v} \\ \hat{\eta}
\end{array}\right)
+
\left(
\begin{array}{c c}
k^2 - \mathcal{D}^2 & 0 \\ 0  & 1  \end{array} \right)^{-1}\left(
\begin{array}{c c}
\mathcal{L}_{OS} & 0 \\ ik_z\overline{u}'  & \mathcal{L}_{SQ}
\end{array}\right)
\left(
\begin{array}{c}
\hat{v} \\ \hat{\eta}
\end{array}\right)
= \boldsymbol{J}\hat{\boldsymbol{f}},
\end{equation}
where the Orr-Sommerfeld (OS) and Squire (SQ) operators are given by
\begin{eqnarray}
\mathcal{L}_{OS} = ik_x\overline{u}(k^2-\mathcal{D}^2) +ik_x\overline{u}'' +\frac{1}{Re_{\tau}}(k^2-\mathcal{D}^2)^2, \\
\mathcal{L}_{SQ} = ik_x\overline{u} + \frac{1}{Re_{\tau}}(k^2-\mathcal{D}^2),
\end{eqnarray}
and
\begin{equation}\label{eq:B_op}
\boldsymbol{J} =
\left(
\begin{array}{c c}
k^2 - \mathcal{D}^2 & 0 \\ 0  & 1  \end{array} \right)^{-1}\left(
\begin{array}{c c c}
-ik_x \mathcal{D} & -k^2 & -ik_z \mathcal{D} \\
  ik_z & 0 & -ik_x
\end{array}\right),
\end{equation}
\begin{equation}
\hat{\boldsymbol{f}}=
\left(
\begin{array}{c}
\hat{f}_u \\ \hat{f}_v \\ \hat{f}_w
\end{array}\right) = -\langle \boldsymbol{u}' \cdot \nabla \boldsymbol{u}' \rangle_{\boldsymbol{k}}.
\end{equation}
Here $\mathcal{D} = \frac{\partial}{\partial{y}}$, $ k^2 = k_x^2 + k_z^2$, and $\langle \hspace{2 mm} \rangle_{\boldsymbol{k}} $ denotes the Fourier component associated with the wavenumber vector $ \boldsymbol{k} = (k_x,k_z,\omega)$. The wall-normal operators are discretized numerically with Chebyshev collocation points using the suite developed by \cite{weideman2000matlab}. We can recast Equation \ref{eq:OS_SQ} into the following input/output form
\begin{equation}\label{eq:IO}
\left(
\begin{array}{c}
\hat{u} \\ \hat{v} \\ \hat{w}
\end{array}\right) = \mathcal{H}(k_x,k_z,\omega)\left(
\begin{array}{c}
\hat{f}_u \\ \hat{f}_v \\ \hat{f}_w
\end{array}\right),
\end{equation}
where the resolvent operator $\mathcal{H}$ is given by
\begin{equation}\label{eq:ResOp}
\mathcal{H}(k_x,k_z,\omega) = \boldsymbol{K}(-i\omega+\boldsymbol{L})^{-1}\boldsymbol{J},
\end{equation}
where
\begin{equation}
\boldsymbol{L} = \boldsymbol{G}^{-1}\mathcal{L},
\end{equation}
\begin{equation}
\boldsymbol{G} = \left(
\begin{array}{c c}
k^2 - \mathcal{D}^2 & 0 \\ 0  & 1  \end{array} \right)  ,
\end{equation}
\begin{equation}\label{eq:OSSQ}
\mathcal{L} = \left(
\begin{array}{c c}
\mathcal{L}_{OS} & 0 \\ ik_z\overline{u}'  & \mathcal{L}_{SQ}
\end{array}\right),
\end{equation}
\begin{equation}
\boldsymbol{K} =
\frac{1}{k^2}\left(
\begin{array}{c c c}
ik_x \mathcal{D} & -ik_z\\ k^2 & 0\\ ik_z \mathcal{D}  & ik_x  \end{array} \right).
\end{equation}
As before, we can decompose the resolvent operator via the SVD as
\begin{equation}
\mathcal{H}(k_x,k_z,\omega) = \boldsymbol{\Psi}(k_x,k_z,\omega)\boldsymbol{\Sigma}(k_x,k_z,\omega)\boldsymbol{\Phi}^{*}(k_x,k_z,\omega)
\end{equation}

The off-diagonal term in $\boldsymbol{L}$ is proportional to the mean shear $ \overline{u}'$ which is maximum at the wall. It remains large in the inner region before it begins to decline in the log region. Similar to the model LNS operator in Section~\ref{sec:amp}, mean shear is the primary source of non-normality leading to significant amplification. Its spatial variation is important since it has been shown \cite{McKeon10} that a critical-layer mechanism tends to localize activity at the wall-normal location where the phase speed of the disturbance is equal to the local mean velocity. This is explored further by considering three particular wavenumber triplets that are representative of the near-wall cycle, a very large-scale motion (VLSM), and a stationary disturbance. The roles of normal and non-normal mechanisms are studied by analyzing the mode shapes of the leading resolvent response modes, the pseudospectrum of the LNS operator, and the resolvent norm compared with the inverse distance between the imaginary axis and the nearest eigenvalue.

\begin{figure}
        \centering
        \subfloat[]{\includegraphics[scale=0.35]{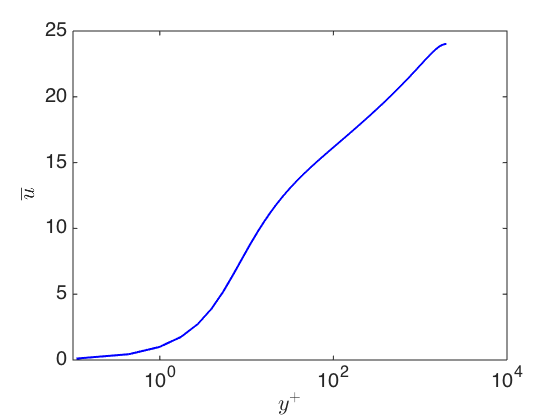}}
        \subfloat[]{\includegraphics[scale=0.35]{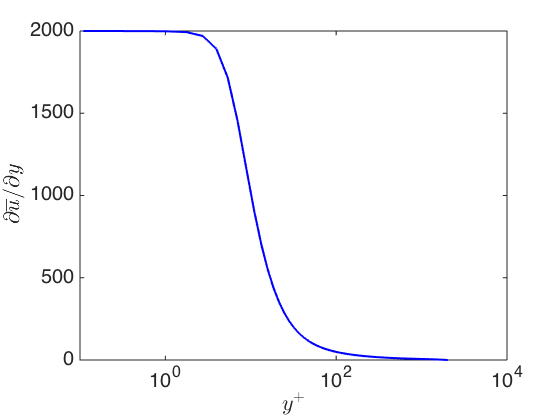}}

\caption{(a) Turbulent mean velocity profile for channel flow at $Re_{\tau} = 2000$ plotted in inner units alongside (b) the mean shear.} \label{fig:turbulent profile}
\end{figure}

\subsection{Near-wall cycle}

\begin{figure}
        \centering
        \subfloat[]{\includegraphics[scale=0.35]{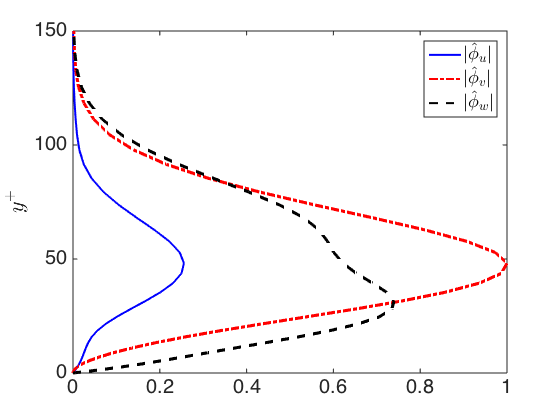}}
        \subfloat[]{\includegraphics[scale=0.35]{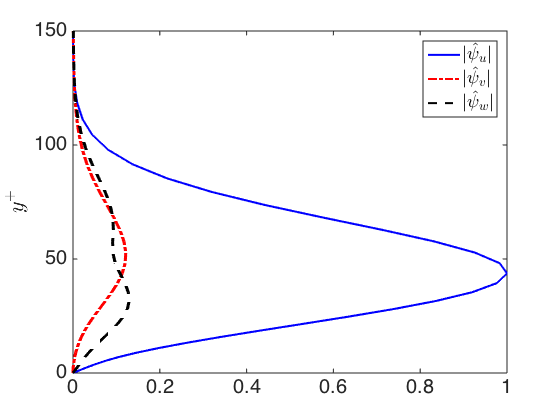}}

\caption{Velocity amplitudes for the optimal forcing mode $\hat{\phi}_1$ in (a) and optimal response mode $\hat{\psi}_1$ in (b) corresponding to the wavenumber triplet of $(k_x,k_z,c^{+}) = (4\pi,40\pi,14)$.}\label{fig:NWC shapes}
\end{figure}

The first wavenumber combination considered is $(k_x,k_z,c^{+}) = (4\pi,40\pi,14)$  which is representative of the near-wall cycle~\cite{McKeon10}. Here the wavespeed is given by $c^{+}=\omega/k_x$. Figure~\ref{fig:turbulent profile} shows that the mean shear is very large at the wall-normal height where the wavespeed matches the local mean, resulting in the off-diagonal terms of the resolvent operator being large. This is similar to the model LNS operator in Equation~\ref{eq:c limit svd} where the influence of non-normality concentrates energy in different velocity components for $\hat{\boldsymbol{\psi}}_1$ and $\hat{\boldsymbol{\phi}}_1$. The optimal resolvent forcing and response modes are plotted in Figure~\ref{fig:NWC shapes} to illustrate that the forcing is primarily concentrated in $v$ and $w$ while the response is mostly in $u$.

\begin{figure}
        \centering
	\includegraphics[scale=0.5]{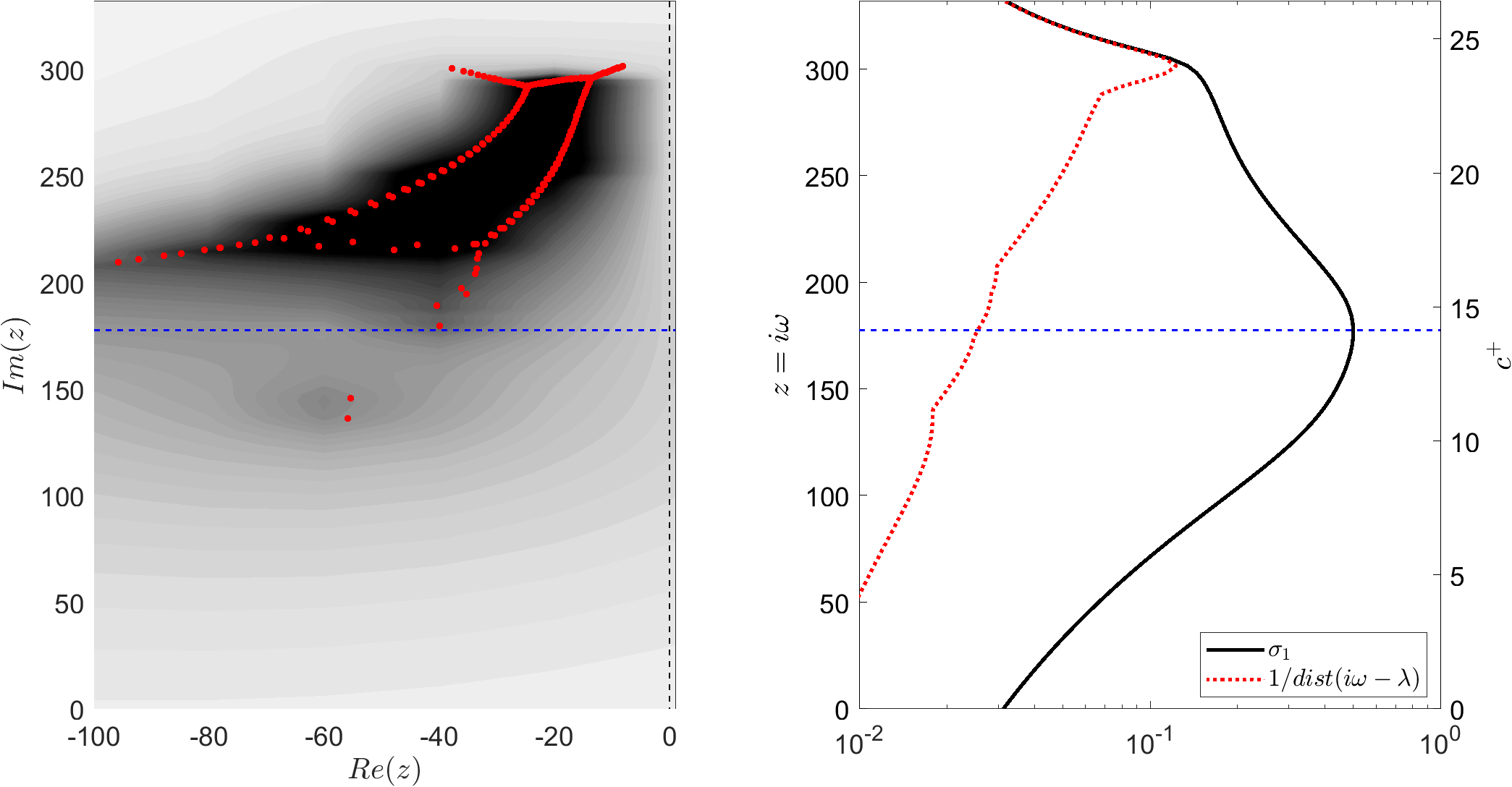}

\caption{The eigenvalues of the operator $\boldsymbol{L}(k_x = 4\pi,k_z = 40\pi)$ in red circles overlaid with contours of the pseudospectrum (left). The resolvent norm is plotted in the solid black line along with the inverse distance from the imaginary axis to the nearest eigenvalue which is the red dotted line (right). The spatial wavenumbers correspond to the near-wall cycle, and the horizontal, dashed blue line represents the $\omega$ with the largest resolvent norm, which corresponds to a phase speed of $c^{+} \approx 14$. }\label{fig:NWC pseudo}
\end{figure}

The strength of mean shear suggests that pseudoresonance is the primary driver of the near-wall cycle mode. The spectrum as well as contours of the pseudospectrum are plotted in Figure~\ref{fig:NWC pseudo} for various $\epsilon$. Figure~\ref{fig:NWC pseudo} also includes the resolvent norm and the contribution from resonance which is quantified by $|i \omega - \lambda|^{-1}$. The ratio of the resolvent norm to the contribution from resonance is 19.6, which is of the same order of magnitude as the value predicted by $| \hat{\boldsymbol{\phi}}_1^{*}\hat{\boldsymbol{\psi}}_1 |^{-1} = 4.81$ (see Table~\ref{tab:mode comparison}). Nevertheless, it is clear from this discrepancy that amplification cannot be attributed to one particular eigenvalue and that there is no eigenvector which is proportional to the resolvent mode. Using the expression $\sigma_1 |i\omega - \lambda|$ to quantify non-normality is problematic since there is no unique $\lambda$ which is responsible for amplification.

\begin{figure}
        \centering
        \includegraphics[scale=0.35]{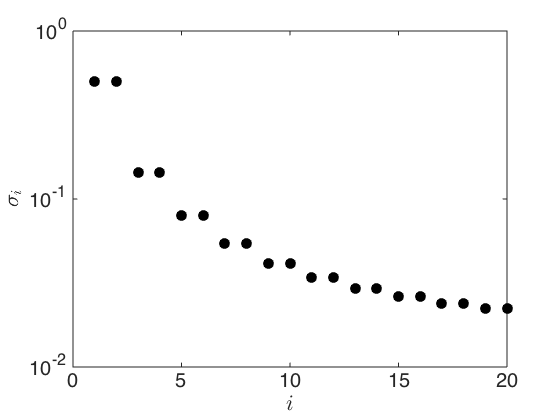}
\caption{First 20 singular values of the resolvent operator for $(k_x,k_z,c^{+}) = (4\pi,40\pi,14)$.}\label{fig:NWC svals}
\end{figure}

The $\omega$ corresponding to $c^{+} = 14$ is indicated by the horizontal, dashed blue line. At this frequency, the resolvent norm is significantly larger than the resonance term suggesting that amplification is due to non-normal mechanisms. This observation is confirmed by the spectrum where the least damped eigenvalues are clustered around higher frequencies and contributes to the resolvent norm for $ \omega > 300$.
It is also worth noting that the eigenvalues are significantly damped, which results in the leading singular values being on the order of unity. While these are not large, the rank-1 approximation is still valid since the first pair of singular values is approximately one order of magnitude larger than the others as seen in Figure~\ref{fig:NWC svals}.

\subsection{VLSM}

\begin{figure}
        \centering
        \subfloat[]{\includegraphics[scale=0.35]{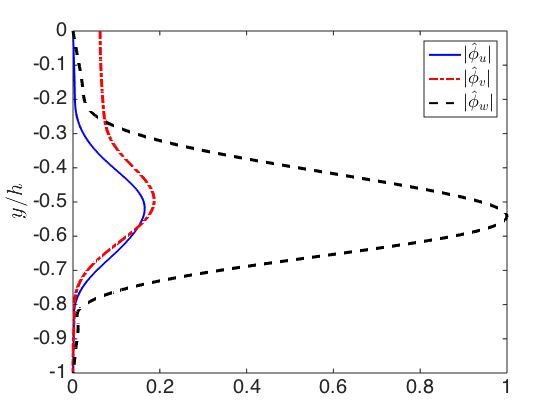}}
        \subfloat[]{\includegraphics[scale=0.35]{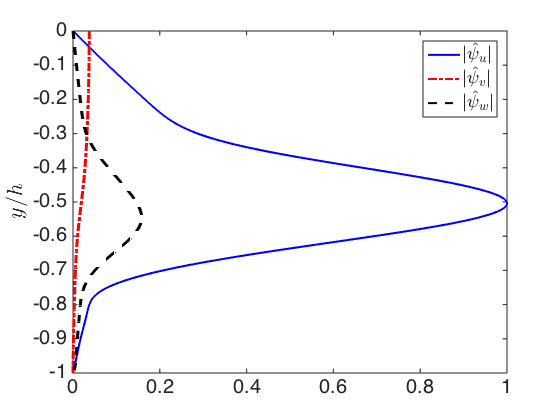}}

\caption{Velocity amplitudes for the optimal forcing mode $\hat{\phi}_1$ in (a) and optimal response mode $\hat{\psi}_1$ in (b) corresponding to the wavenumber triplet of $(k_x,k_z,c^{+}) = (\pi/9,2\pi /3,22)$.}\label{fig:VLSM shapes}
\end{figure}

\begin{figure}
        \centering
	\includegraphics[scale=0.5]{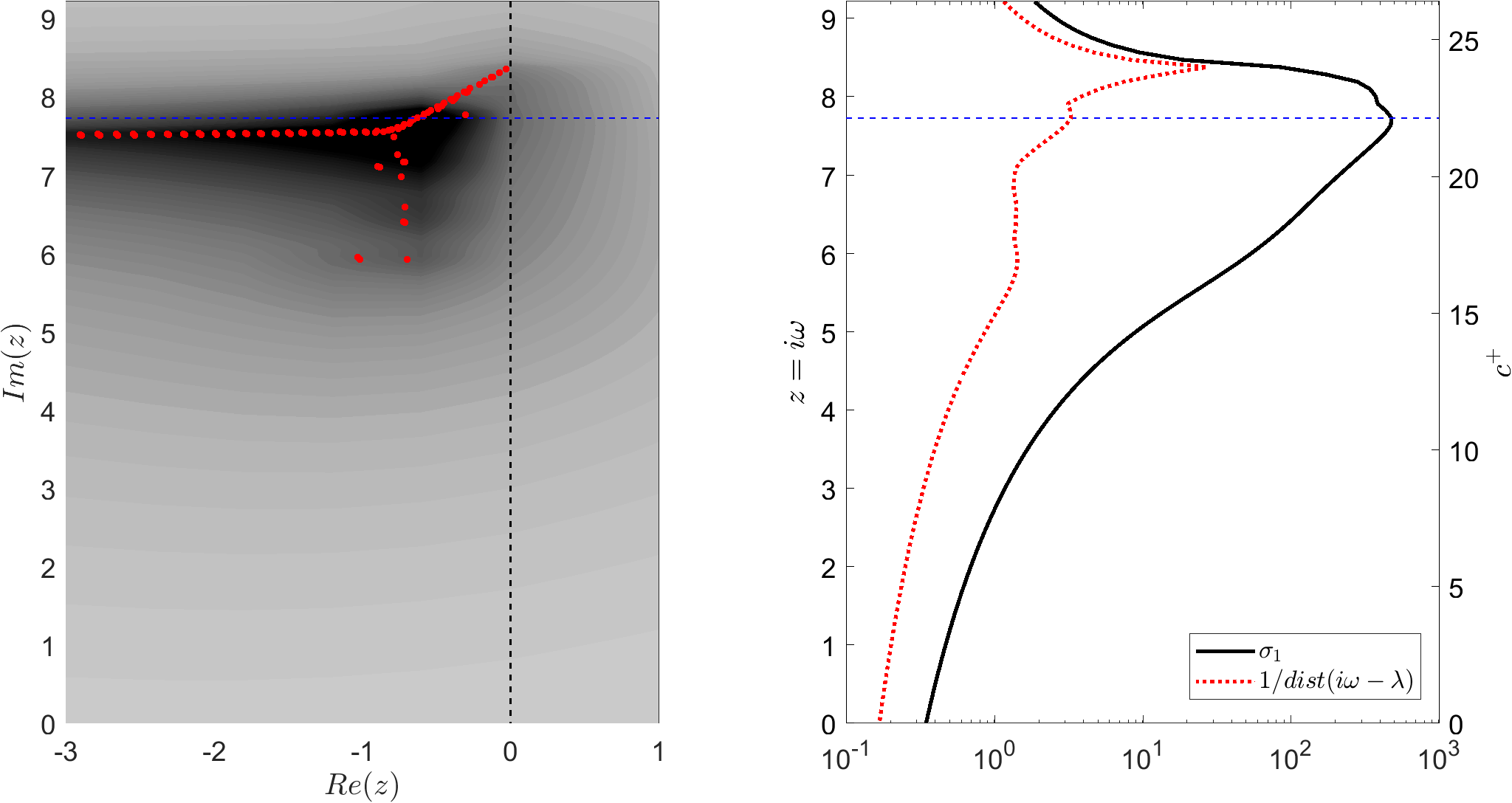}

\caption{The eigenvalues of the operator $\boldsymbol{L}(k_x = \pi/9,k_z = 2\pi/3)$ in red circles overlaid with contours of the pseudospectrum (left). The resolvent norm is plotted in the solid black line along with the inverse distance from the imaginary axis to the nearest eigenvalue which is the red dotted line (right).  The spatial wavenumbers correspond to the VLSM mode, and the horizontal, dashed blue line represents the $\omega$ with the largest resolvent norm, which corresponds to a phase speed of $c^{+} \approx 22$.}\label{fig:VLSM pseudo}
\end{figure}

Further from the wall, the mean shear drops several orders of magnitude (see Figure~\ref{fig:turbulent profile}) and the effect of viscosity decreases. As a result, amplification becomes a mix of both normal and non-normal effects. To reinforce this point, a wavenumber triplet which is representative of a VLSM is considered. The mode shapes for $(k_x,k_z,c^{+}) = (\pi/9,2\pi /3,22)$ are plotted in Figure~\ref{fig:VLSM shapes}.
The forcing is dominated by the $w$-component while the response is dominated by the $u$-component. The $v$-component of the forcing, notably, is less significant than the near-wall cycle mode implying that the role of lift-up is not as pronounced for this mode. The spectrum associated with the streamwise and spanwise wavenumbers of the VLSM is plotted along with the pseudospectrum of the LNS operator in Figure~\ref{fig:VLSM pseudo}.

\begin{figure}
        \centering
        \includegraphics[scale=0.35]{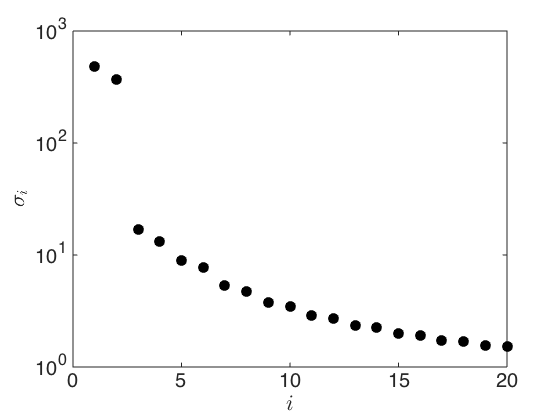}

\caption{First 20 singular values of the resolvent operator for $(k_x,k_z,c^{+}) = (\pi/9,2\pi /3,22)$.}\label{fig:VLSM svals}
\end{figure}

The results are drastically different from the near-wall cycle case as the resonant forcing of eigenvalues is greater than one so amplification is due to both terms on the right-hand side of Equation~\ref{eq:resolvent eigenvalue}. Notably, the spectrum in Figure~\ref{fig:VLSM pseudo} resembles that of the base flow at $Re = 10,000$ for $k_x = 1$, $k_z = 0$ as there are three distinct branches. As observed by~\citet{Reddy93} and~\citet{Schmid01}, the eigenvalues at the intersection of the branches are the most sensitive to perturbations and result in very large non-normal amplification. The product $\sigma_1 | i \omega - \lambda | = 146$ while $\|\hat{\boldsymbol{\phi}}_1^{*}\hat{\boldsymbol{\psi}}_1 \|^{-1}$ = 34.9 (see Table~\ref{tab:mode comparison}), suggesting that the nonorthogonality of many eigenfunctions leads to high pseudoresonance. Similar to the near-wall cycle case, there are no eigenvalues which exactly match the wave speed associated with the VLSM. The eigenvalue close to the dotted blue line in Figure~\ref{fig:VLSM pseudo}, however, does seem to impact the resolvent norm which has an extra bump near its maximum value. This is also reflected in the red dotted line since the eigenvalue protrudes from the distinct Y-shape of the spectrum. The maximum singular values are on the order of $10^3$ and the resolvent operator is low-rank as seen in Figure~\ref{fig:VLSM svals}.

\subsection{Stationary disturbances}

\begin{figure}
        \centering
        \subfloat[]{\includegraphics[scale=0.35]{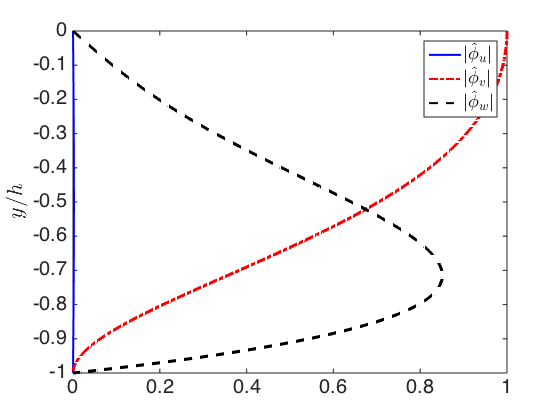}}
        \subfloat[]{\includegraphics[scale=0.35]{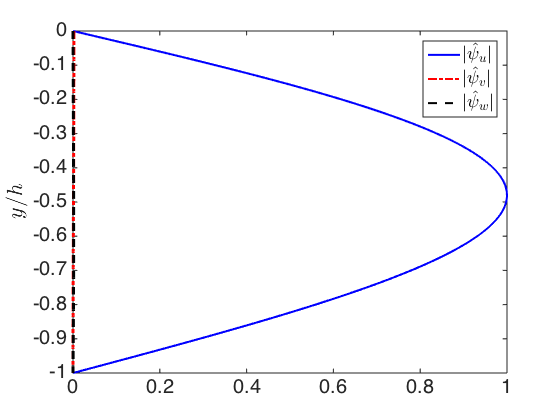}}

\caption{Velocity amplitudes for the optimal forcing mode $\hat{\phi}_1$ in (a) and optimal response mode $\hat{\psi}_1$ in (b) corresponding to the wavenumber triplet of $(k_x,k_z,\omega) = (0,2\pi /3,0)$.}\label{fig:stationary shapes}
\end{figure}

Finally, we consider stationary disturbances which tend to be the most amplified by the resolvent operator with singular values exceeding $10^4$. The specific wavenumber triplet selected for this study is $(k_x,k_z,\omega) = (0,2\pi/3,0 )$. The roots behind such large amplification can be traced back to the model operator in Equation~\ref{eq:c limit svd}. In this example, the $\partial \overline{u}/\partial y \to \infty$ resulting in a low-rank system which concentrated all the forcing energy in the second velocity component and the response energy in the first velocity component. When $k_x = \omega = 0$, all of the diagonal terms of the resolvent become order $\epsilon$ small since imaginary terms are eliminated and $\mathcal{D}$ scales with $Re^{-1}$. Thus, when the LNS operator is inverted, the determinant, which is the product of the diagonal terms, is very small. The energy for the forcing is almost totally in the wall-normal and spanwise components as seen in Figure~\ref{fig:stationary shapes} while the response is almost totally in the streamwise component.

\begin{figure}
        \centering
	\includegraphics[scale=0.4]{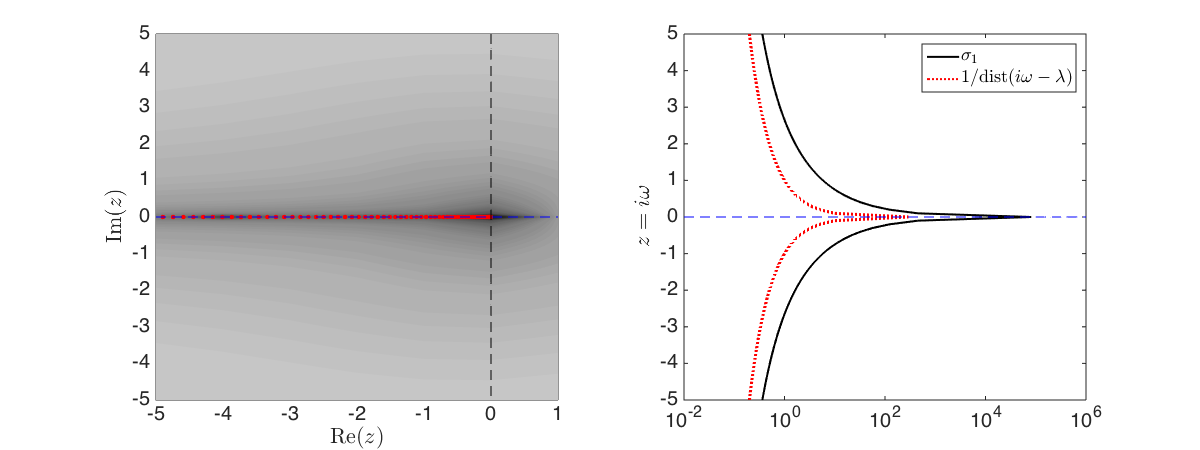}

\caption{The eigenvalues of the operator $\boldsymbol{L}(k_x = 0,k_z = 2\pi /3)$ in red circles overlaid with contours of the pseudospectrum (left). The resolvent norm is plotted in the solid black line along with the inverse distance from the imaginary axis to the nearest eigenvalue which is the red dotted line (right). The horizontal, dashed blue line represents the $\omega$ corresponding to the phase speed of $c^{+} = 0$.}\label{fig:stationary pseudo}
\end{figure}

\begin{figure}
        \centering
        \includegraphics[scale=0.35]{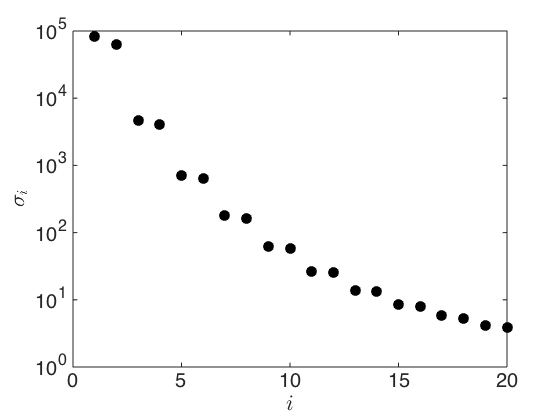}

\caption{First 20 singular values of the resolvent operator for $(k_x,k_z,\omega) = (0,2\pi /3,0)$.}\label{fig:stationary svals}
\end{figure}

Similar to the near-wall cycle and VLSM modes, the spectrum and contours of the pseudospectrum are presented in Figure~\ref{fig:stationary pseudo} alongside the resolvent norm and contribution from resonance. All of the eigenvalues are real since the imaginary terms are eliminated from the resolvent operator when $k_x = 0$. Another implication of eliminating mean flow advection, as mentioned by~\cite{Hack17}, is that the Orr mechanism is absent from this mode.
Stationary disturbances are highly amplified and the singular values are plotted in Figure~\ref{fig:stationary svals}. The rank-1 approximation is quite applicable for this particular mode as the leading pair of singular values is on the order of $10^5$. Moreover, the contribution from non-normality is well approximated by $| \hat{\boldsymbol{\phi}}_1^{*}\hat{\boldsymbol{\psi}}_1 | ^{-1} = 250$ which agrees quite well with $\sigma_1 | i\omega - \lambda |$ = 278. Such agreement can be attributed to the eigenvalue closest to the imaginary axis which has an imaginary component that agrees with the most amplified frequency. Unlike the cylinder case where there exists a convective-type non-normality, the $k_x = 0$ modes are an example of the component-type non-normality and so $| \hat{\boldsymbol{\phi}}_1^{*}\hat{\boldsymbol{\psi}}_1 |$ is small since the velocity for the forcing mode is almost completely concentrated in the wall-normal plane while the velocity for the response mode is nearly all in the streamwise direction.

\subsection{Predicting low-rank nature of resolvent}

We briefly return to the discussion of predicting when the resolvent operator may be low-rank. When the LNS operator is inverted, the terms which were originally small along the diagonal become large leading to a large resolvent norm. Marginally stable eigenvalues or stationary disturbances have large amplification since either one term along the diagonal is small when forcing at a resonant frequency or all the terms along the diagonal are small since $k_x = \omega = 0$. The system is low-rank roughly speaking if the largest entry of the resolvent operator is appreciably larger than the other terms. This can happen when the off-diagonal are terms are large with respect to the diagonal as is the case for the near-wall cycle and stationary disturbance modes or when one term along the diagonal is small as is the case for the cylinder. If the frequency of the forcing excites two eigenvalues then two diagonal terms of the resolvent operator are large and the system is not rank-1 even though there might be significant amplification (see Appendix~\ref{sec:rank1}).

\subsection{Influence of spatial wavenumber and wave speed}

Based on the findings of this study and observations from~\cite{Bourg12}, it is possible to categorize amplification mechanisms in wall-bounded turbulence as either normal or non-normal depending on the wavenumber vector $\boldsymbol{k}$ selected. When $k_x$ is small, the influence of both mean flow advection and viscosity is diminished resulting in a non-normal system where there is high amplification. Low-order modes (those corresponding to the largest singular values), experience high amplification due to both normality and non-normality. Higher-order modes also experience amplification due to non-normality. At higher $k_x$, only low-order modes are amplified as long as they are localized near the critical layer. Higher-order modes experience low amplification which is proportional to viscosity. The wall-normal height, furthermore, has implications on the type of amplification as it influences the choice of wave speed $c$, or $\omega$, as well as the influence of mean shear $\partial \overline{u}/\partial y$. Closer to the wall, the mean shear is highest while in the log region, mean shear still plays an important role resulting in preferential amplification of long streamwise structures.

\begin{table}
\begin{center}
\begin{tabular}{ r | c | c | c | c | c |}
\multicolumn{1}{r}{}
 & \multicolumn{1}{c}{$\boldsymbol{k}$ or $\omega$}
 & \multicolumn{1}{c}{$\sigma_1$}
 & \multicolumn{1}{c}{$|i\omega-\lambda|^{-1}$}
 & \multicolumn{1}{c}{$\sigma_1|i\omega-\lambda|$}
 & \multicolumn{1}{c}{$|\hat{\boldsymbol{\phi}}_1^{*}\hat{\boldsymbol{\psi}}_1 |^{-1}$} \\
\cline{2-6}
& & & & & \\
Cylinder Base~  & ~$\omega = 0.743$~ & ~5.78e04~ & ~729~ & ~79.3~ & ~79.4~ \\
& & & & & \\
\cline{2-6}
& & & & & \\
Cylinder Mean~ & ~$\omega = 1.02$~ & ~1.65e04~ & ~570~ & ~28.9~ & ~26.9~ \\
& & & & & \\
\cline{2-6}
& & & & & \\
Near-wall cycle~ & ~$(k_x,k_z,c^{+}) = (4\pi, 40\pi,14)$ & ~0.502~ & ~2.56e-02~ & ~19.6~ & ~4.81~ \\
& & & & & \\
\cline{2-6}
& & & & & \\
VLSM~ & ~$(k_x,k_z,c^{+}) = (\pi/9, 2\pi/3, 22) $~ & ~479~ & ~3.28~ & ~146~ & ~34.9~ \\
& & & & & \\
\cline{2-6}
& & & & & \\
Stationary Disturbance~ & ~$(k_x,k_z,\omega) = (0, 2\pi/3, 0)$~ & ~8.14e04~ & ~293~  & ~278~ & ~250~ \\
& & & & & \\
\cline{2-6}
\end{tabular}
\end{center}
\caption{Quantification of non-normality for the most amplified modes in cylinder and turbulent channel flows.} \label{tab:mode comparison}
\end{table}

\section{Closing remarks}
\label{sec:conclusion}

\subsection{Conclusions}

The preceding development and example operators provide a context in which to interpret the recent findings concerning analysis of the resolvent in the literature. Amplification quantified by the resolvent norm incorporates both normal and non-normal origins. In the limiting case of a normal operator, the condition number $\kappa = 1$ and resonance due to the eigenvalue is captured.  For a non-normal operator, $\kappa \gg 1 $ and pseudoresonant effects can lead to large amplification. The nature of the non-normality in the resolvent contains information about the transfer mechanisms from inputs to outputs and may impact the distribution of energy among different velocity components. The inverse of the inner product $| \hat{\boldsymbol{\phi}}_1^*\hat{\boldsymbol{\psi}}_1|$, which is unity for a purely normal mechanism and zero for a purely non-normal mechanism, quantifies the degree to which the amplification is non-normal. In the presence of mean shear, there is always some degree of non-normality so a mechanism can be considered normal in character if $| \hat{\boldsymbol{\phi}}_1^*\hat{\boldsymbol{\psi}}_1|$ is close to unity.

The model resolvent operator is also non-self-adjoint unless there is no mean shear, the eigenvalues are real, and the disturbances are stationary ($\omega = 0$). If the eigenvalues are purely imaginary and there is no spatial dependence, the resolvent forcing and response modes are $90^{\circ}$ out of phase. In a spatially-developing flow, this translates to the spatial support of the forcing mode being upstream that of the response mode. It is also possible to observe the Orr mechanism which aligns the resolvent response and forcing modes with and against the mean shear, respectively. If there is reverse flow, the modes overlap and form a wavemaker which is approximately coincident with the region of the flow that is absolute unstable. Non-normality can still be quantified via $| \hat{\boldsymbol{\phi}}_1^*\hat{\boldsymbol{\psi}}_1 |^{-1}$, which tends to infinity for a purely convective-type instability mechanism. It is concluded that mode shapes and locations can be generalized depending on the strength of lift-up and mean flow advection.

These findings are applicable to both base and mean velocity profiles, but there is an important distinction. For a base flow, the input forcing must be provided by an external source. Therefore, the most amplified structure is of interest as it is likely to be present in the flow when it becomes unsteady. For the cylinder flow~\citep{Barkley06} and thermo-solutal convection~\citep{Turton15}, the operator is normal in character so the most amplified structure corresponds to an eigenvalue crossing the imaginary axis. This is not the case for non-normal operators~\citep[e.g.][]{Jovanovic05} as the highest amplification occurs for modes which do not correspond to an eigenvalue. For a mean flow, the nonlinear term is the source of intrinsic forcing so the goal is to predict the structure of the unsteady flow~\citep{Taira17}. In many cases, the most amplified response is normal in character and accounts for a significant amount of the kinetic energy of the velocity fluctuations~\citep[e.g.][]{McKeon10, Beneddine16}. There are circumstances, however, where it does not give the complete picture as the rank-1 approximation is no longer valid and suboptimal modes need to be considered.

Expressing the resolvent operator as the outer products of eigenmodes clarifies the connection between stability and resolvent analysis. Furthermore, it becomes apparent that mean stability analysis is successful when the amplification is due to forcing at a frequency in the vicinity of a unique eigenvalue. In such circumstances, the forward and adjoint stability modes are proportional to the optimal forcing and response modes at that frequency. Both sets of modes form a rank-1 approximation, which fails if the eigenvalues are clustered together, of the resolvent operator. Cylinder flow is chosen to demonstrate the similarity of stability and resolvent modes when resonance outweighs pseudoresonance at a particular frequency. In both the base flow and mean flow cases, amplification occurs at a single frequency corresponding to the imaginary part of the least stable eigenvalue. The resolvent modes approximate the region of the flow which is absolutely unstable by identifying the wavemaker. There is no wavemaker at very low Reynolds numbers when the flow is only convectively unstable. The cylinder exemplifies a convective-type non-normality where mean flow advection separates the spatial support of the forcing mode to be upstream of the response mode as long as $\overline{u} > 0$. Non-normality quantified by $| \hat{\boldsymbol{\phi}}_1^*\hat{\boldsymbol{\psi}}_1 |^{-1}$ agrees well with $\sigma_1 |i\omega - \lambda| $ for both the base and mean flows.

It is important to note that only normal mechanisms are active in cylinder flow whereas in more complicated flows such as wall-bounded turbulence, both normal and non-normal mechanisms are relevant. Three different wavenumber triplets, representative of the near-wall cycle, VLSM's, and stationary disturbances, highlight the competing influences of viscous dissipation, mean flow advection, and mean shear on not only the most amplified modes but also the spectrum and pseudospectrum. The importance of each term depends significantly on the wall-normal height where the perturbations are localized. In the inner region where there is very high mean shear and viscosity is most important, amplification is primarily due to pseudoresonant mechanisms. Forcing in the spanwise/wall-normal plane leads to a large response in the streamwise direction as seen for the near-wall cycle mode. The eigenvalues are highly damped yet the resolvent norm is on the order of unity due to the high sensitivity of the spectrum to perturbations. In the log region, mean shear and consequently lift-up are weaker yet the declining importance of viscosity results in eigenvalues which are closer to the imaginary axis. The most amplified disturbance corresponding to the wavespeed of the VLSM, nevertheless, is primarily due to pseudoresonance. Consequently, there is poorer agreement between non-normality quantified by $| \hat{\boldsymbol{\phi}}_1^*\hat{\boldsymbol{\psi}}_1 |^{-1}$ and $\sigma_1 |i\omega - \lambda| $ as it is clear that amplification can no longer be attributed to a single eigenvalue in the spectrum. Mean flow advection results in the Orr mechanism \citep[see][]{McKeon17} and hence less overlap between the forcing and response modes.

Stationary disturbances are the globally most amplified disturbances by effectively leveraging mean shear. The perturbation energy is almost exclusively concentrated in the $v-$ and $w$-components of the forcing mode and in the $u$-component of the response mode. Assuming streamwise constant disturbances eliminates the mean flow advection term from the resolvent operator and hence suppresses the Orr mechanism. All of the non-normality, consequently, can be classified as a component-type non-normality, in contrast to the cylinder flow, and the eigenvalues of the LNS operator are real. Non-normality quantified by $| \hat{\boldsymbol{\phi}}_1^*\hat{\boldsymbol{\psi}}_1 |^{-1}$ agrees well with $\sigma_1 |i\omega - \lambda| $ since amplification can be attributed to the eigenvalue closest to the imaginary axis. For a generic wavenumber triplet, the applicability of the rank-1 approximation can be approximated by the ratio of the largest term in the resolvent operator to the other entries. Finally, the distribution of energy among various velocity components may be useful when considering how the nonlinear term, which can be computed from resolvent response modes \citep{McKeon13}, projects onto the optimal resolvent forcing modes.

\subsection{Reduced-order modeling and control}

In the pseudo-resonance limit case discussed in Section~\ref{sec:amp}, it was demonstrated that a single resolvent forcing/response pair could characterize the most amplified disturbance. The same could not be said of the eigenvectors which were nearly parallel to one another. In reduced-order modeling applications, the resolvent modes could be a superior basis to the stability modes, since fewer are needed to represent the most important behavior of the flow. In the context of control, where reduced-order models play a key role in rendering the problem computationally tractable, stability modes have been shown by~\citet{Barbagallo09} to be less effective at representing the input-output behavior of the system. While their study did not consider resolvent modes as an expansion basis for model reduction, their capability of capturing the intrinsic dynamics (stability-related phenomena) as well as non-modal growth could make them an ideal candidate. Resolvent modes can also provide relevant information in regard to where to place sensors and actuators.

The resolvent response modes are outputs which are highly amplified by the linear dynamics of the NSE, and so sensors could be placed where these are likely to be strong. The resolvent forcing modes are the `trigger' or input which leads to high amplification, and so the actuators could be placed to manipulate the flow in such a way that suppresses these disturbances.

Note, however, that in cases where there is large spatial separation between resolvent forcing and response modes, it is possible that improved performance could be attained by sensing and actuating within a wavemaker region \citep[e.g.][]{Chen11, Giannetti07}, which, as discussed in Section \ref{sec:cylinder}, may also be estimated from resolvent analysis.  While the resolvent decomposition shows potential for control applications \citep{Luhar14}, further refinements could seek to balance the observability and controllability of the reduced-order model \citep{Moore81,Rowley05}, subject to known sensor and actuator locations, and information about the nature of the nonlinear forcing.

\section{Acknowledgments}
\label{sec:acknowledgments}

The work in this study has been financially supported by a National Science Foundation Graduate Fellowship, AFOSR under grant number FA 9550-16-1-0361 and ARO under grant number W911NF-17-1-0306. The authors would like to thank Denis Sipp for providing the resolvent code and Andres Goza who assisted setting it up on a cluster. Finally, the authors would like to acknowledge Theresa Saxton-Fox and Ryan McMullen for useful feedback on the manuscript. 

\appendix

\section{Rank-1 example}
\label{sec:rank1}

\begin{figure}
        \centering
        \subfloat[]{\includegraphics[scale=0.35]{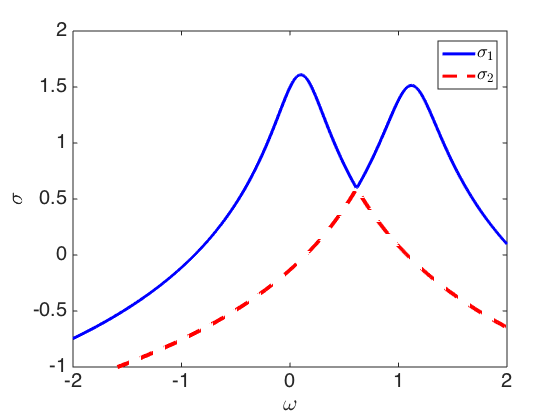}}
        \subfloat[]{\includegraphics[scale=0.35]{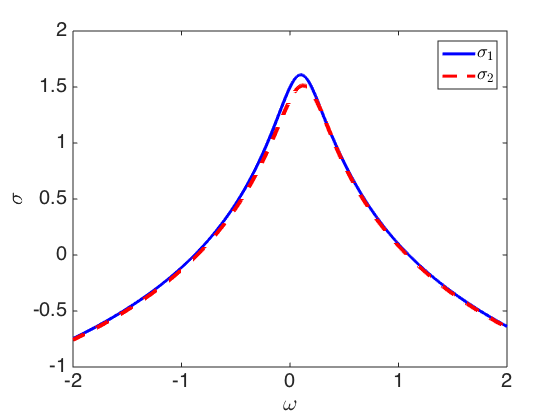}}
\caption{The singular values as a function of the forcing frequency for the operators $\boldsymbol{T}_1$ (left) and $\boldsymbol{T}_2$ (right).} \label{fig:rank1}
\end{figure}

An important result of this paper is that the rank-1 approximation fails when there is not sufficient separation of eigenvalues at the frequency of interest. The two operators $\boldsymbol{T}_1$ and $\boldsymbol{T}_2$ in Equation~\ref{eq:T1 and T2},  which are nearly identical except for the frequency of the second eigenvalue, illustrate this point. In the case of $\boldsymbol{T}_2$, substituting the eigenvalues into Equation~\ref{eq:dyad H} results in two terms of roughly equal importance leading to the failure of the rank-1 approximation as seen in Figure~\ref{fig:rank1} where the singular values of the resolvent operator for both $\boldsymbol{T}_1$ and $\boldsymbol{T}_2$ are plotted. The rank-1 approximation works better for $\boldsymbol{T}_1$ since there is a clear separation of singular values at the resonant frequencies.

\begin{equation}
\boldsymbol{T}_1 = \left( \begin{array}{cc} -0.2+0.1i & 0 \\ 0 & -0.22+\boxed{1.12i} \end{array} \right), ~~~ \boldsymbol{T}_2 = \left( \begin{array}{cc}  -0.2+0.1i & 0 \\ 0 & -0.22+\boxed{0.12i} \end{array} \right).
\label{eq:T1 and T2}
\end{equation}

\bibliography{cylinder}% Produces the bibliography via BibTeX.

\end{document}